\documentclass[11pt,a4paper]{article}
\pdfoutput=1
\usepackage{jheppub}
\usepackage{amsmath,amsfonts,amssymb, mathtools, mathrsfs}
\usepackage{hyperref, csquotes}
\usepackage{graphicx, color}
\usepackage{tikz,pgf,tikz-cd,float}
\usepackage{physics,tensor,subcaption}

\newcommand{\R}{\mathbb{R}}

\newcommand{\e}{\mathrm{e}}

\newcommand{\Sreg}{S_{\mathrm{R}}}

\newcommand{\GN}{G_{\mathrm{N}}}

\begin{document}

\title{Cosmology in Lorentzian Regge calculus: causality violations, massless scalar field and discrete dynamics}

\author[a,b,c]{Alexander F. Jercher,}
\emailAdd{alexander.jercher@uni-jena.de}

\author[a]{Sebastian Steinhaus}
\emailAdd{sebastian.steinhaus@uni-jena.de}

\affiliation[a]{Theoretisch-Physikalisches Institut, Friedrich-Schiller-Universit\"{a}t Jena\\ Max-Wien-Platz 1, 07743 Jena, Germany, EU}
\affiliation[b]{Arnold Sommerfeld Center for Theoretical Physics,\\ Ludwig-Maximilians-Universit\"at München \\ Theresienstrasse 37, 80333 M\"unchen, Germany, EU}
\affiliation[c]{Munich Center for Quantum Science and Technology (MCQST),\\ Schellingstr. 4, 80799 M\"unchen, Germany, EU}

\date{\today}

\begin{abstract}
{
We develop a model of spatially flat, homogeneous and isotropic cosmology in Lorentzian Regge calculus, employing 4-dimensional Lorentzian frusta as building blocks. By examining the causal structure of the discrete spacetimes obtained by gluing such 4-frusta in spatial and temporal direction, we find causality violations if the sub-cells connecting spatial slices are spacelike. A Wick rotation to the Euclidean theory can be defined globally by a complexification of the variables and an analytic continuation of the action. Introducing a discrete free massless scalar field, we study its equations of motion and show that it evolves monotonically. Furthermore, in a continuum limit, we obtain the equations of a homogeneous scalar field on a spatially flat Friedmann background. Vacuum solutions to the causally regular Regge equations are static and flat and show a restoration of time reparametrisation invariance. In the presence of a scalar field, the height of a frustum is a dynamical variable that has a solution if causality violations are absent and if an inequality relating geometric and matter boundary data is satisfied. Edge lengths of cubes evolve monotonically, yielding a contracting or an expanding branch of the universe. In a small deficit angle expansion, the system can be deparametrised via the scalar field and a continuum limit of the discrete theory can be defined which we show to yield the relational Friedmann equation. These properties are obstructed if higher orders of the deficit angle are taken into account. Our results suggest that the inclusion of timelike sub-cells is necessary for a causally regular classical evolution in this symmetry restricted setting. Ultimately, this works serves as a basis for forthcoming investigations on the cosmological path integral within the framework of effective spin foams.
}
\end{abstract}

\maketitle

\section{Introduction}

Deriving quantum cosmology from full quantum gravity is an important research goal for multiple reasons: the large degree of symmetry greatly reduces the complexity of the considered quantum gravity model. As a result, analytical studies and numerical simulations are more accessible. Moreover, this simplified setting permits tackling of conceptual questions, such as matter coupling, causal structure of (quantum) spacetime, recovery of semi-classical physics and, in approaches that feature discrete variables, remnants of continuum symmetries and the definition of a continuum limit. On the phenomenological side, quantum effects are expected to become relevant for physics of the early Universe, in particular close to the Big Bang. These effects might leave traces detectable in cosmological observations such as that of the CMB spectrum~\cite{Planck:2018vyg}, therefore offering a promising testing ground for quantum gravity theories.

Quantum gravity approaches with discrete degrees of freedom, such as spin foam models~\cite{Perez:2013uz}, loop quantum gravity (LQG)~\cite{Ashtekar:2021kfp} and group field theories (GFTs)~\cite{Oriti:2006ts,Freidel:2005qe}, face two major challenges in this endeavour. First, the identification of macroscopic cosmological variables from the microscopic ones. This involves either a coarse-graining procedure as in GFT condensate cosmology~\cite{Marchetti:2020umh,Oriti:2016qtz,Gielen:2016dss,Jercher:2021bie} or a symmetry reduction. The latter can be performed either before quantising, i.e.~on the classical configuration space, as in loop quantum cosmology~\cite{Agullo:2016tjh} or after quantising, i.e.~on the quantum configuration space, as e.g. in~\cite{Bahr:2017bn,Bianchi:2010ej}. The second challenge is to define the dynamics of cosmological observables in the absence of a background manifold. Evolution of observables can only be described with respect to other dynamical degrees of freedom, leading to a relational formulation~\cite{Rovelli:1990ph,Rovelli:2001bz,Dittrich:2005kc,Hoehn:2019fsy}.

In spin foams a cosmological subsector is identified by symmetry reducing to those quantum geometric variables which capture the desired spatially homogeneous and isotropic dynamics. In earlier works of spin foam cosmology~\cite{Bianchi:2010ej,Bianchi:2011ym,Rennert:2013pfa,Sarno:2018ses} the cosmological dynamics were examined in particular transition amplitudes of so-called dipole states and petal states in the full EPRL spin foam model~\cite{Engle:2007em,Engle:2008fj}. Recently, investigations were conducted to define a Hartle-Hawking no boundary wavefunction by computing the transition from nothing to an equilateral spatial triangulation in the EPRL model~\cite{Gozzini:2019nbo,Frisoni:2022urv,Frisoni:2023lvb}. Other examples of symmetry reduced spin foam models, not necessarily tailored to cosmology, are given by the cuboid~\cite{Bahr:2015gxa,Bahr:2016dl,Bahr:2017klw}, frusta~\cite{Bahr:2017bn,Bahr:2018ewi,Bahr:2018gwf,Jercher:2023rno} and parallel-epiped models~\cite{Assanioussi:2020fml}.  Due to the numerical challenges accompanying the evaluation of symmetry reduced quantum amplitudes for large spins~\cite{Allen:2022unb}, resorting to semi-classical amplitudes appeared fruitful in advancing explicit computations. Following this line of research, effective spin foams~\cite{Asante:2021zzh,Asante:2020iwm,Asante:2020qpa} have been applied to the cosmological setting in~\cite{Dittrich:2023rcr,Dittrich:2021gww,Asante:2021phx} extending earlier works of quantum Regge cosmology~\cite{CorreiadaSilva:1999cg,CorreiadaSilva:1999es,Hartle:1985wr,Hartle:1986up,Liu:2015gpa}. As the semi-classical geometry, one frequently considers an equilateral triangulation of the $3$-sphere which describe positively curved spatial slices and which evolves into a $3$-sphere of different size, describing the evolution of the scale factor.

Recent studies of effective spin foams (and associated classical Regge calculus)~\cite{Asante:2021zzh,Dittrich:2021gww,Asante:2021phx,Dittrich:2023rcr} have revealed several intriguing physical features already present in these simplified cosmological models. The triangulated spacetimes are piece-wise Lorentzian, i.e. its building blocks are pieces of flat Minkowski spacetime glued together. This allows for a plethora of configurations and different cases, since sub-simplices can be either space- or timelike. 
In turn, this can lead to causally irregular configurations in the form of more or less than two light cones located at spacelike sub-simplices, some of which can be interpreted as topology change.
In the cosmological setting mentioned above, one typically obtains causally regular configurations if the edges connecting the spatial $3$-spheres are timelike. Such a case is similar to triangulations encountered in Causal Dynamical Triangulations (CDT)~\cite{Ambjorn:2012jv,Loll:2019rdj}. However, if all building blocks are spacelike causally irregular configurations generically appear. 
To find a causally regular subsector of the full quantum theory therefore suggests the investigation of causally extended spin foam models such as the Conrady-Hnybida (CH) extension of the Lorentzian EPRL model~\cite{Conrady:2010kc,Conrady:2010vx} or the completion of the Barrett-Crane model~\cite{Jercher:2022mky}. Unfortunately these models are not as computationally mature as their purely spacelike formulation and open questions regarding their asymptotics remain (see~\cite{Kaminski:2017eew,Liu:2018gfc,Simao:2021qno} for the asymptotics of the EPRL-CH model).

In this article we are proposing a different spatial triangulation to describe cosmology. Instead of a constantly curved space, we propose flat space described by a cubulation with toroidal topology. These tori are connected in spacetime by so-called frusta, higher-dimensional trapezoids, which interpolate between cubes of different size. In fact, we are adapting the proposal from the Euclidean EPRL model~\cite{Bahr:2017bn} to Lorentzian Regge calculus, where the 3-frusta can be either space- or timelike. Similar to the triangulated $3$-spheres, we will show that causally irregular structures can occur if some building blocks become spacelike; in most such cases the associated Regge action is complex and classical solutions do not exist. In fact, in vacuum spacetimes, the only permitted solutions are static spacetimes, where the spatial slices are connected by timelike edges. These solutions are flat and thus exhibit a reparametrisation invariance~\cite{Rocek:1981ama,Rocek:1982tj}: the time elapsed between two slices is arbitrary, which can be related to a change of lapse function. In such a setting, taking the continuum limit is trivial, both in temporal and spatial directions.\footnote{In the $3$-sphere model, a spatial continuum limit cannot be straightforwardly defined without deviating from homogeneity. Only three equilateral triangulations exist consisting of five, $16$ or $600$ tetrahedra.}

The second direction we are exploring here is coupling a massless scalar field to the classical Regge action, following the suggestion by Hamber~\cite{Hamber:1993gn}. Our goal is to investigate whether this scalar field can be used to deparametrise the theory, i.e. express the cosmological dynamics relationally in terms of the scalar field and thus in a diffeomorphism invariant manner. In our case this would imply that the dependence on the height variables, essentially the proper time elapsed between two spatial slices, drops out and is replaced by differences of the scalar field. If this is the case, the foliation we have chosen would be exactly the one singled out by the scalar field. This insight should however be taken with a grain of salt as we have made several simplifying assumptions. In this article, we will explicitly solve the classical dynamics of the scalar field and observe a strictly monotonic behaviour (depending on the boundary values of the scalar field and the discrete geometry).

The question whether the theory can be deparametrised is more subtle: as soon as the scalar field boundary data are not equal, we will show that the Regge geometry is no longer flat and thus loses the reparametrisation invariance. Hence, the height variables become dynamical degrees of freedom and are uniquely fixed by the equations of motion. While we may still attempt to express the dynamics in terms of the scalar field, an explicit dependence on the height variables remains. The situation is further complicated by the transcendental nature of the Regge equations. To better understand how the symmetry is broken (and eventually restored in a continuum limit), we expand the equations of motion in a Taylor series of small deficit angles. To lowest order in the deficit angles, the system is reparametrisation invariant and the dependence on the height variables can be replaced by expressing the dynamics with respect to scalar field differences. This breaks down at higher orders of the deficit angles, which is not surprising. Thus, we expect to restore the symmetry in a continuum limit, in which the deficit angles become small and eventually vanish. We study this continuum limit and recover the classical solution of spatially flat cosmology with non-vanishing massless scalar field.

Moreover, in most cases adding a scalar field leads to a regular causal structure. The solutions to the Regge equations are not flat, where the spatial slices are connected by timelike edges. Hence the Regge action is purely real and the geometry causally regular. The situation changes if we change the sign in front of the scalar field action. Then, no classical solutions can be found: in the causally regular sector, the Regge equations have no solutions. In the irregular sector, the Regge action is complex and the equations of motion for real and imaginary part are not simultaneously solved for real length variables.

In short, the purpose of this article is to define a spatially flat cosmological subsector in Lorentzian Regge calculus coupled to a massless scalar field. The thorough classical analysis of the system we are about to present is to prepare the eventual examination of the quantum system as a path integral in effective spin foams. Our article is organised as follows: in Sec.~\ref{sec:Spatially flat cosmology in Regge calculus} we introduce the triangulation and Lorentzian Regge calculus, discuss irregular causal structures and define a Wick rotation. Sec.~\ref{sec:Minimally coupled scalar field} introduces the minimally coupled scalar field in the discrete, its equations of motion, continuum limit and Wick rotation as well. In Sec.~\ref{sec:Regge equations and deparametrisation} we study the Regge equations of motion for multiple time steps in vacuum and with matter, we linearise around small deficit angles and discuss deparametrisation as well as causality violations and their relation to matter. We close in Sec.~\ref{sec:Discussion and conclusion} with a discussion and conclusion.

\section{Spatially flat cosmology in Regge calculus}\label{sec:Spatially flat cosmology in Regge calculus}

Regge calculus~\cite{Regge:1961ct} is a discrete gravitational theory, introduced for simplicial manifolds $\Delta$ with dynamical variables being the length of edges $\{l_e\}$. In four dimensions, to which we restrict to in the remainder, a simplicial manifold consists of $4$-simplices $\sigma\in\Delta$ which are internally flat, i.e. they can be embedded into Minkowski space $\R^{1,3}$. Curvature is located at the triangles $t\in\Delta$ and captured by deficit angles $\delta_t(\{l_e\})$, which lie in a two-dimensional space orthogonal to $t$. In the Lorentzian setting, this space is two-dimensional Minkowski space $\R^{1,1}$ or two-dimensional Euclidean space $\R^2$, depending on the triangle $t$ being spacelike or timelike\footnote{In this work we neglect the presence of lightlike edges, triangles and tetrahedra. In fact, the Regge action in $d$ dimensions is insensitive to the causal character of $(d-k)$-dimensional sub-cells for $k>2$ including edges in the $4$-dimensional case. Furthermore, since the volumes of $(d-2)$ cells enter the Regge action linearly, contributions from lightlike hinges (with vanishing volume) are zero. Finally, $(d-1)$-dimensional null cells require the definition of angles between lightlike vectors for which we refer the reader to~\cite{Sorkin:2019llw}. In this work, we choose boundary data such that $(d-1)$-dimensional cells are not null.}, respectively. To define Lorentzian angles we follow in this work the conventions of~\cite{Sorkin:2019llw,Asante:2021zzh} which ensure additivity of angles as in Euclidean space. The deficit angles can then be expressed as
\begin{equation}\label{eq:general deficit angles}
\delta_t^{\mathrm{E}} = 2\pi-\sum_{\sigma\supset t}\psi_{\sigma,t}^{\mathrm{E}},\qquad \delta_t^{\mathrm{L},\pm} = \mp i 2\pi  -\sum_{\sigma\supset t}\psi_{\sigma,t}^{\mathrm{L},\pm},
\end{equation}
where the $\psi_{\sigma,t}$ are dihedral angles, i.e. the angles between normal vectors $N_\tau$ of tetrahedra $\tau\supset t$ within the $4$-simplex $\sigma$. Here, the \enquote{$\pm$} indicates a choice of sign which becomes important if the imaginary parts of the Lorentzian dihedral angles do not sum up to $\pm i 2\pi$. As we will discuss in more detail in Sec.~\ref{sec:causal regularity}, such configurations are interpreted as causally irregular, representing for instance spatial topology change~\cite{Sorkin:2019llw,Asante:2021zzh,Asante:2021phx}.

The bulk action of Lorentzian Regge calculus is given by~\cite{Sorkin:2019llw}
\begin{equation}\label{eq:general Regge action}
\Sreg[\{l_e\}] = \sum_{t\in\Delta:\text{ tl}}\abs{A_t(\{l_e\})}\delta_t^{\mathrm{E}}(\{l_e\})+\sum_{t\in\Delta:\text{ sl}}\abs{A_t(\{l_e\})}\delta_t^{\mathrm{L},\pm}(\{l_e\}),
\end{equation}
where $\abs{A_t}$ is the absolute value of the area associated to the triangle $t$. In the presence of a boundary $\partial\Delta$, the Regge action attains boundary terms such that overall additivity is ensured. The equations of motion are given by
\begin{equation}
\pdv{\Sreg}{l_e} = \sum_t \pdv{\abs{A_t}}{l_e}\delta_t = 0,
\end{equation}
where the Schl\"{a}fli identity has been imposed, i.e. $\sum_t \abs{A_t}\pdv{\delta_t}{l_e} = 0$. 

There exist several related formulations of Regge calculus in terms of different variables~\cite{Barrett:1994ba,Dittrich:2008hg}, most notably area Regge calculus~\cite{Asante:2018wqy,Makela:2000ej,Barrett:1997tx}, where the dihedral angles are a function of areas rather than lengths. Since $10$ lengths uniquely characterise a $4$-simplex while $10$ areas do not and since a triangulation contains in general more triangles than edges, area variables do in general lead to different equations of motion than length variables. As shown in~\cite{Barrett:1997tx,Dittrich:2008va,Asante:2021zzh}, area variables need to be further constrained to yield the dynamical equations of length Regge calculus. This idea lies at the heart of effective spin foams~\cite{Asante:2021phx,Asante:2020iwm,Asante:2020qpa}, where a gravitational path integral is implemented for area Regge calculus, where the area-length constraints are imposed ad-hoc via Gaussians. For simple, symmetry reduced systems, such as the cosmological models in~\cite{Asante:2021phx,Dittrich:2023rcr,Dittrich:2021gww} as well as the model we construct in the following, the relation of area and length is globally invertible and the constraints mentioned above trivialise. 

\begin{figure}
    \centering
    \includegraphics[width=0.7\textwidth]{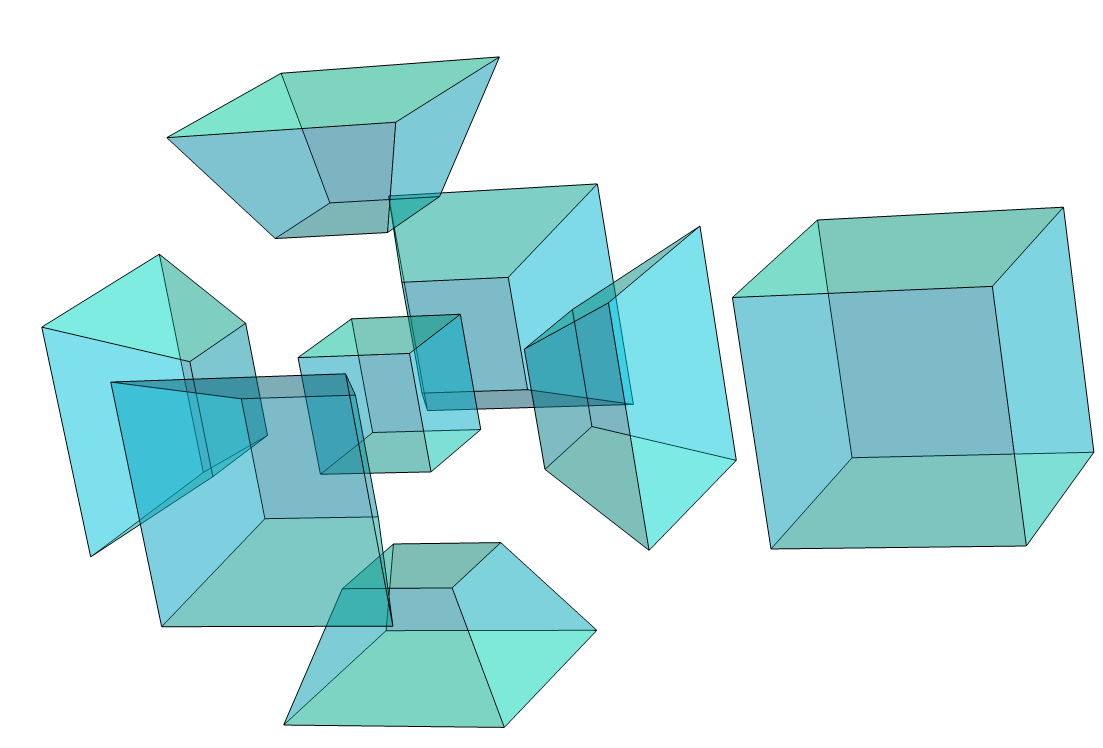}
    \caption{The boundary of a $4$-frustum, consisting of two $3$-cubes and six equal boundary $3$-frusta, connecting the cubes. Not explicitly visualised here, the two $3$-cubes lie in different spacelike hypersurfaces and have a timelike separation.}
    \label{fig:4frustumbdry}
\end{figure}

\begin{figure}
    \centering
    \includegraphics[width=0.6\textwidth]{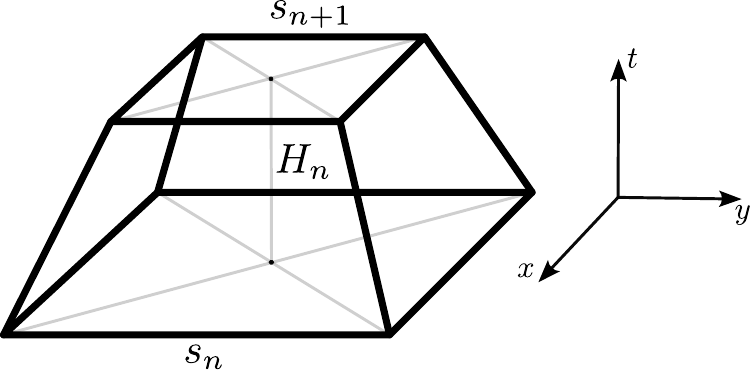}
    \caption{Depiction of a $4$-frustum with one spatial dimension suppressed. Spacelike edges of $3$-cubes, lying in different spacelike hypersurfaces, have a length of $s_n$ and $s_{n+1}$, respectively. The height $H_n$ of a $4$-frustum is defined as the distance between the midpoints of the $3$-cubes.}
    \label{fig:4frustum3d}
\end{figure}

Goal of this section is to set up a Regge action for spatially flat and isotropic cosmology. To that end, we consider at the level of combinatorics a special class of simplicial manifolds which consist of triangulated $4$-dimensional cuboids\footnote{Examples for triangulations of a $4$-dimensional cube can be found in~\cite{ScottMara1976,Dittrich:2021kzs,Dittrich:2022yoo}.}. We assume that each $4$-dimensional triangulated polytope admits an analogue of foliation into spacelike hypersurfaces, that is, there are two $3$-dimensional triangulated polytopes at different instances of time, connected by six $3$-dimensional triangulated polytopes. Spatial homogeneity translates to the requirement that every $3$-polytope in a fixed spacelike slice is the same. Isotropy is imposed by requiring the six faces of spacelike $3$-polytopes to be equal. Finally, imposing spatial flatness fixes the $4$-dimensional building block to be a triangulated Lorentzian $4$-frustum, being the $4$-dimensional generalisation of a trapezoid. A Lorentzian $4$-frustum consists of two spacelike $3$-cubes in different spacelike hypersurfaces connected by six boundary $3$-frusta, see Fig.~\ref{fig:4frustumbdry} for a visualisation. The requirements of spatial flatness, homogeneity and isotropy are so restrictive that the entire geometry of a triangulated $4$-frustum can be captured by the length of the $3$-cubes, $s_0$ and $s_1$, and its height $H$, defined as the distance between the midpoints of the two $3$-cubes (see Fig.~\ref{fig:4frustum3d} for a $3$-dimensional depiction).\footnote{Originally, the height $H$ is not part of the variables as it does not refer to a length of any of the edges. However, as we see from Eq.~\eqref{eq:edge length}, the length of the edge connecting cubes of different slices can be expressed in terms of the height and hence, these variables can be used interchangeably. Using $H$ as a dynamical variable allows for a characterisation of the geometry which is closer to the continuum with lapse function $N$.} 

Deficit angles at subdividing triangles vanish as the inside of the $4$-frustum and its contained sub-cells are constrained to be flat.  Hence, the Regge action associated to this type of restricted triangulated geometries effectively reduces to a generalised Regge action of $4$-frusta geometries, the $2$-dimensional building blocks of which are squares and trapezoids.  Henceforth, we focus on $4$-frusta geometries and neglect any further flat triangulation.

Gluing $4$-frusta along boundary $3$-frusta in spatial direction and along boundary $3$-cubes in temporal direction, one obtains an extended discrete cosmological spacetime with cubulated slices labelled by $n\in\{0,...,M\}$. We say that a $4$-frustum lies in a \enquote{slab} between slices $n$ and $n+1$.

In the following three sections, we characterise the geometry of Lorentzian $4$-frusta more explicitly to ultimately set up the action for spatially flat cosmology in Lorentzian Regge calculus in Sec.~\ref{sec:Lorentzian Regge action for spatially flat cosmology}. 

\subsection{Lorentzian frustum}\label{sec:Lorentzian frustum}

In a Lorentzian setting, the edges, trapezoids and $3$-frusta connecting slices $n$ and $n+1$ can be either of spacelike or timelike character, reflected in the sign of the squared length, area and $3$-volume, respectively.\footnote{We use the sign convention $(-,+,+,+)$ for the Minkowski metric $\eta$. Hence, vectors with $\eta(v,v)>0$ $(<0)$ are spacelike (timelike).} In the following, we express these geometric quantities as functions of $(s_n,s_{n+1},H_n)$ and show under which conditions the building blocks are spacelike or timelike.

\paragraph{Edges.} Edges contained in trapezoids which thus connect cubes that lie in distinct slices $n$ and $n+1$ have a squared edge length given by
\begin{equation}\label{eq:edge length}
l_n^2 = 3\left(\frac{s_n-s_{n+1}}{2}\right)^2-H_n^2,
\end{equation}
and therefore we have that:
\begin{center}
\begin{tabular}{c c c c}
  edge is timelike if   & $l_n^2 < 0$ & $\Leftrightarrow$ & $H_n^2 > \frac{3}{4}(s_n-s_{n+1})^2$, \\[7pt]
  edge is spacelike if  & $l_n^2 > 0$ & $\Leftrightarrow$ & $H_n^2 < \frac{3}{4}(s_n-s_{n+1})^2$.
\end{tabular}
\end{center}

\paragraph{Trapezoids.} The squared area $k_n^2$ of a trapezoid  is given by
\begin{equation}
k_n^2 = \left(\frac{s_n+s_{n+1}}{2}\right)^2\left[\left(\frac{s_n-s_{n+1}}{\sqrt{2}}\right)^2-H_n^2\right],
\end{equation}
from which we extract that:
\begin{center}
\begin{tabular}{c c c c}
  trapezoid is timelike if   & $k_n^2 < 0$ & $\Leftrightarrow$ & $H_n^2 > \frac{1}{2}(s_n-s_{n+1})^2$, \\[7pt]
  trapezoid is spacelike if  & $k_n^2 > 0$ & $\Leftrightarrow$ & $H_n^2 < \frac{1}{2}(s_n-s_{n+1})^2$.
\end{tabular}
\end{center}

\paragraph{$\mathbf{3}$-frusta.} To determine the signature and the $3$-volume of $3$-frusta, we consider first the squared height of the $3$-frustum,
\begin{equation}
h_n^2 = \left(\frac{s_n-s_{n+1}}{2}\right)^2-H_n^2.
\end{equation}
Then, the signature of the $3$-frustum is determined by the signature of its height:
\begin{center}
\begin{tabular}{c c c c}
    $3$-frustum is timelike if & $h_n^2 < 0$ & $\Leftrightarrow$ & $H_n^2 > \frac{1}{4}(s_n-s_{n+1})^2$,\\[7pt]
    $3$-frustum is spacelike if & $h_n^2 > 0$ & $\Leftrightarrow$ & $H_n^2 < \frac{1}{4}(s_n-s_{n+1})^2$.
\end{tabular}
\end{center}
Using the squared height of the $3$-frustum, $h_n^2$, one can express the squared $3$-volume as
\begin{equation}
v_n^2 = \frac{(s_n^2+s_ns_{n+1}+s_{n+1}^2)^2}{9}h_n^2 =  \frac{(s_n^2+s_ns_{n+1}+s_{n+1}^2)^2}{9}\left[\left(\frac{s_n-s_{n+1}}{2}\right)^2-H_n^2\right]. 
\end{equation}

\paragraph{$\mathbf{4}$-frusta.} Lastly, the squared $4$-volume of a $4$-frustum is given by
\begin{equation}
V_n^2 = \frac{(s_n^2+s_{n+1}^2)^2(s_n+s_{n+1})^2}{16}H_n^2
\end{equation}
As a building block of top dimension, the Lorentzian $4$-frustum is constrained to have a positive $4$-volume, yielding the condition of a positive squared $4$-height, i.e. $H_n^2 > 0$. If $H_n^2 < 0$, the $4$-frustum can only be embedded in $4$-dimensional Euclidean space.

\begin{figure}
    \centering
    \includegraphics[width=\textwidth]{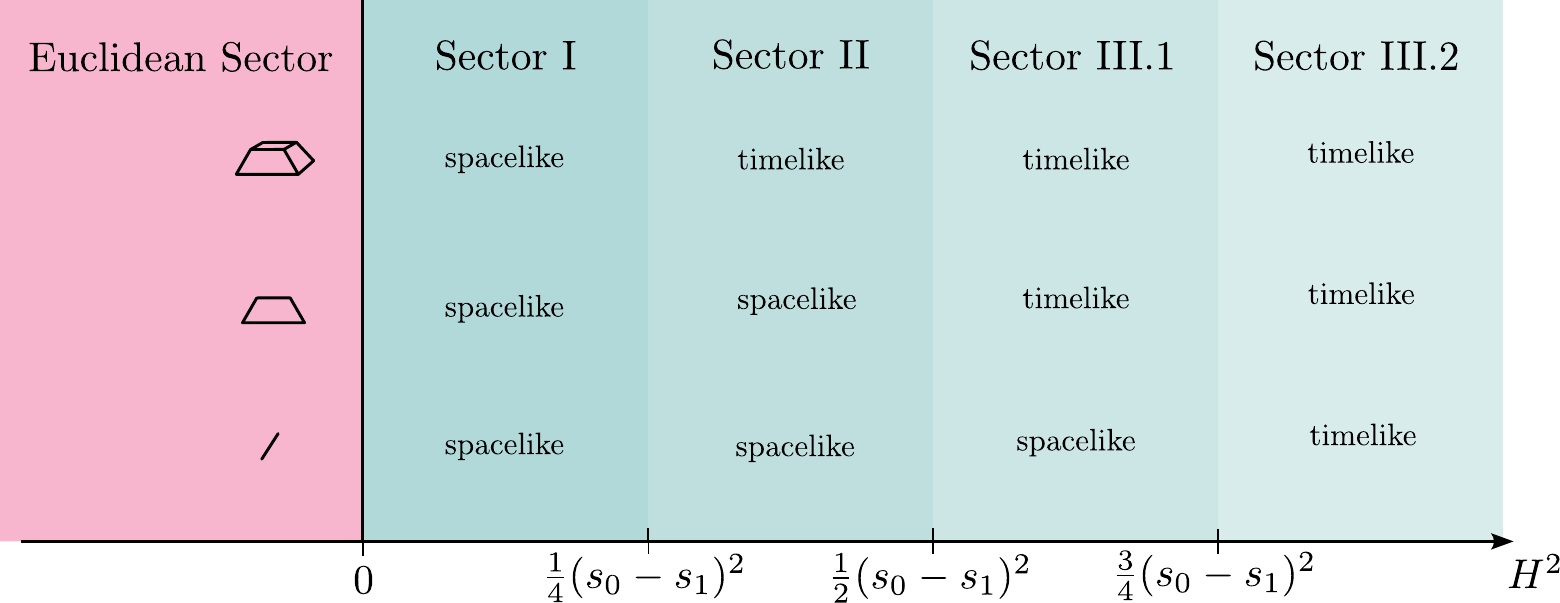}
    \caption{Sectors, determined by the relation of height $H$ and difference in spacelike edge length $(s_0-s_1)$, in which the sub-cells of Lorentzian $4$-frusta take different signature.}
    \label{fig:regions}
\end{figure}

Summarizing, the causal character of the lower-dimensional sub-cells is determined by the difference of spacelike edge length $s_n$ and $s_{n+1}$ relative to the $4$-dimensional height $H_n$. Depicted in Fig.~\ref{fig:regions}, these relations are consistent with the fact that spacelike cells can only contain spacelike sub-cells while timelike sub-cells need always to be contained in timelike cells. The region of $H_n^2 > 0$ corresponds to the Lorentzian sector of the theory while $H_n^2 < 0$ is associated to the Euclidean sector. As we argue in Sec.~\ref{sec:Wick rotation}, $H_n$ is naturally associated to the lapse function $N$ which, in the simple context of homogeneous cosmology, relates the Lorentzian and Euclidean sector via an analytical continuation, commonly referred to as Wick rotation. 

\subsection{Lorentzian angles}\label{eq:Lorentzian angles}

Following the introduction of Regge calculus at the beginning of this section, dihedral and deficit angles form a crucial ingredient for setting up the Lorentzian Regge action. In the restricted setting we consider here, there are two types of dihedral angles, associated either to spacelike squares or to trapezoids connecting slices. Crucially, the explicit expressions of these angles depend on the causal characters of sub-cells of the $4$-frustum and thus on the relative size of the variables $(s_n,s_{n+1},H_n)$. In the following, we provide a list of the dihedral angles for the various cases, using the definitions of Lorentzian angles as given in~\cite{Asante:2021zzh}. For a more detailed derivation, see Appendix~\ref{sec:Lorentzian dihedral angles of 4-frusta}.

\paragraph{Dihedral angles at squares.} Cubes lie in spacelike hypersurfaces such that the squares contained in it must also be spacelike. As a consequence, the space orthogonal to a square is two-dimensional Minkowski space $\R^{1,1}$. In that space, the dihedral angle between the $3$-cube and the $3$-frustum meeting at that square is given by the Lorentzian angle between the respective projected normal vectors. While the normal vector to a $3$-cube is always timelike the signature of the vector normal to the $3$-frustum can be either spacelike or timelike, opposite to the signature of the $3$-frustum. Within a $4$-frustum, we refer to the dihedral angle located at the (past) $n$-th, respectively the (future) $(n+1)$-th slice as $\varphi_{nn+1}$ and $\varphi_{n+1n}$. Their explicit definitions in terms of the variables $(s_n,s_{n+1},H_n)$ are given by
\begin{equation}\label{eq:phi nn+1}
\varphi_{nn+1} = 
\begin{cases}
-\cosh^{-1}\left(\frac{s_{n+1}-s_n}{\sqrt{(s_n-s_{n+1})^2-4H_n^2}}\right) &\mathrm{if}\quad H_n^2 <\frac{1}{4}(s_n-s_{n+1})^2\quad\mathrm{and}\quad s_n<s_{n+1},\\[10pt]
\cosh^{-1}\left(\frac{s_n - s_{n+1}}{\sqrt{(s_n-s_{n+1})^2-4H_n^2}}\right) \mp i\pi &\mathrm{if}\quad H_n^2 <\frac{1}{4}(s_n-s_{n+1})^2\quad\mathrm{and}\quad s_n>s_{n+1},\\[10pt]
\sinh^{-1}\left(\frac{s_n-s_{n+1}}{\sqrt{4H_n^2-(s_n-s_{n+1})^2}}\right)\mp i\frac{\pi}{2} &\mathrm{if}\quad H_n^2 > \frac{1}{4}(s_n-s_{n+1})^2,
\end{cases}
\end{equation}
and
\begin{equation}\label{eq:phi n+1n}
\varphi_{n+1n} = 
\begin{cases}
-\cosh^{-1}\left(\frac{s_n-s_{n+1}}{\sqrt{(s_n-s_{n+1})^2-4H_n^2}}\right) &\mathrm{if}\quad H_n^2 <\frac{1}{4}(s_n-s_{n+1})^2\quad\mathrm{and}\quad s_n > s_{n+1},\\[10pt]
\cosh^{-1}\left(\frac{s_{n+1}-s_n}{\sqrt{(s_n-s_{n+1})^2-4H_n^2}}\right) \mp i\pi &\mathrm{if}\quad H_n^2 <\frac{1}{4}(s_n-s_{n+1})^2\quad\mathrm{and}\quad s_n < s_{n+1},\\[10pt]
\sinh^{-1}\left(\frac{s_{n+1}-s_n}{\sqrt{4H_n^2-(s_n-s_{n+1})^2}}\right) \mp i\frac{\pi}{2} &\mathrm{if}\quad H_n^2 > \frac{1}{4}(s_n-s_{n+1})^2.
\end{cases}
\end{equation}
We observe that the real parts of the angles $\varphi_{nn+1}$ and $\varphi_{n+1n}$ are related by a minus sign. Furthermore, we notice that for a timelike frustum, the dihedral angles $\varphi$ do not depend on the sign of $s_n-s_{n+1}$. That is because the associated normal vector is spacelike and therefore insensitive to the time orientation.

\paragraph{Dihedral angles at trapezoids.} The dihedral between two $3$-frusta is located at a trapezoid, which can be either spacelike or timelike.

To a spacelike trapezoid, implying the inequality $H_n^2<\frac{1}{2}(s_n-s_{n+1})^2$, we associate the Lorentzian dihedral angle
\begin{align}\label{eq:theta L}
\theta_n^{\mathrm{L}} = 
\begin{cases}
-\cosh^{-1}\left(\frac{(s_n-s_{n+1})^2}{(s_n-s_{n+1})^2-4H_n^2}\right)&\mathrm{if}\quad H_n^2<\frac{1}{4}(s_n-s_{n+1})^2,\\[10pt]
-\cosh^{-1}\left(\frac{(s_n-s_{n+1})^2}{4H_n^2-(s_n-s_{n+1})^2}\right) \mp i\pi &\mathrm{if}\quad H_n^2 > \frac{1}{4}(s_n-s_{n+1})^2.
\end{cases}
\end{align}

To a timelike trapezoid, i.e. $H_n^2 > \frac{1}{2}(s_n-s_{n+1})^2$, we associate the Euclidean angle 
\begin{equation}\label{eq:theta E}
\theta_n^{\mathrm{E}} = \cos^{-1}\left(\frac{(s_n-s_{n+1})^2}{4H_n^2-(s_n-s_{n+1})^2}\right).
\end{equation}

This completes the introduction of geometrical quantities of a single $4$-frustum lying in a slab bounded by spatial slices $n$ and $n+1$.

\subsection{Lorentzian Regge action for spatially flat cosmology}\label{sec:Lorentzian Regge action for spatially flat cosmology}

\begin{figure}
    \centering
    \includegraphics[width=0.6\textwidth]{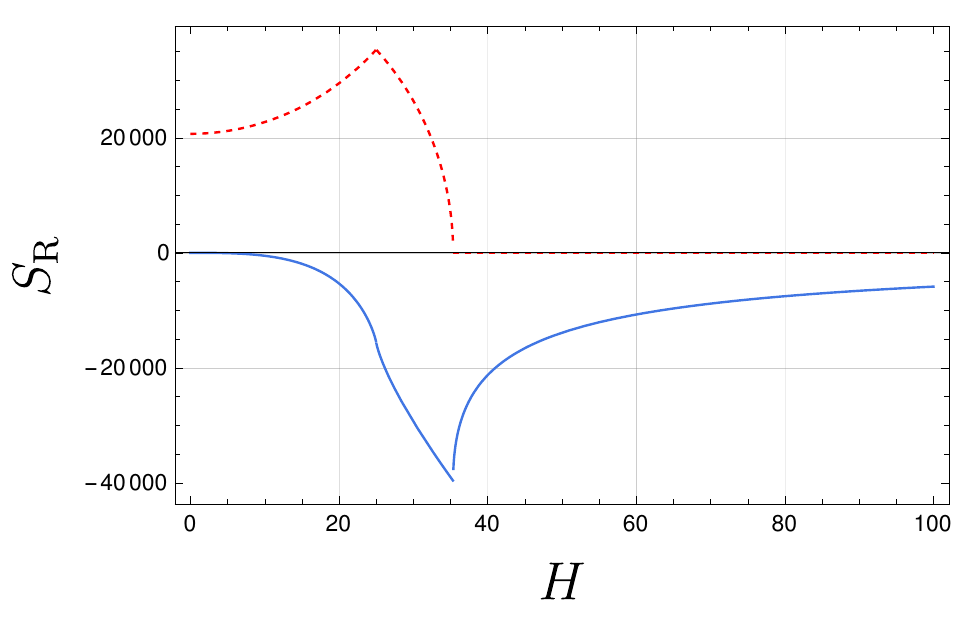}
    \caption{The real (solid blue) and imaginary (dashed red) parts of the Lorentzian Regge action of a single slice for boundary data $s_1 = 50$, $s_2 = 100$ plotted against the height $H$. The kinks in the Regge action stem from the Heaviside $\Theta$-functions that characterise the different causal sectors. Notice, that at the kinks, the Regge action is not differentiable and these points therefore do not correspond to non-trivial solutions of the equations of motion with respect to the variable $H$.}
    \label{fig:VacRegge}
\end{figure}

To construct the Lorentzian Regge action for spatially flat homogeneous cosmology with $M$ spatial slices, we consider first the pure boundary Regge action $S_{\mathrm{R}}^{(n)}$ of a single $4$-frustum bounded by the slices $n$ and $n+1$. This is guided by the condition that the resulting action is additive for larger complexes, i.e. if $4$-frusta are glued together both in timelike and spacelike direction. The underlying $2$-complex defining the combinatorics of the model is hypercubic such that every square or trapezoid is shared by four $4$-frusta. As a result, the exterior boundary deficit angle for a single frustum is defined as
\begin{align}
 \delta^{\mathrm{E}} &= \frac{\pi}{2} - \psi^{\mathrm{E}},\\[7pt]
 \delta^{\mathrm{L},\pm} &= \mp i\frac{\pi}{2} - \psi^{\mathrm{L},\pm},
\end{align}
for angles in Euclidean space, $\psi^{\mathrm{E}}$, and Lorentzian space, $\psi^{\mathrm{L},\pm}$, respectively. Since the single $4$-frustum consists of six squares in slice $n$, twelve trapezoids and six squares in slice $n+1$, spatial homogeneity then implies that the Regge action is given by
\begin{equation}\label{eq:slice Regge action}
\begin{aligned}
S_{\mathrm{R}}^{(n)}[s_n,s_{n+1},H_n] &= 6s_n^2\left(\mp i\frac{\pi}{2}-\varphi_{nn+1}\right)+6s_{n+1}^2\left(\mp i\frac{\pi}{2}-\varphi_{n+1n}\right)\\[7pt]
&+ 12\abs{k_n}\left[\Theta_{\mathrm{sl}}\left(\mp i\frac{\pi}{2}-\theta_n^{\mathrm{L}}\right) + \Theta_{\mathrm{tl}}\left(\frac{\pi}{2} - \theta_n^{\mathrm{E}}\right)\right],
\end{aligned}
\end{equation}
where the $\Theta_{\mathrm{sl,tl}}$ are Heaviside functions that impose the trapezoid to be spacelike or timelike, respectively. An exemplary plot of the Regge action for $M = 1$ is given in Fig.~\ref{fig:VacRegge}.

By construction, the $4$-frustum action of Eq.~\eqref{eq:slice Regge action} is additive such that on a hypercubic complex with $L$ vertices in spatial direction and $M+1$ spatial slices, the action is given by
\begin{equation}\label{eq:Regge action}
S_{\mathrm{R}} = L\sum_{n=0}^{M-1}S_{\mathrm{R}}^{(n)}[s_n,s_{n+1},H_n].
\end{equation}
Additivity of the action per slab implies that the equations of motion for a given geometric quantity, being either $s_n$ or $H_n$, will only depend on neighbouring geometric labels. We discuss the consequences of this slicing structure for the dynamics of the model in more detail in Sec.~\ref{sec:Regge equations and deparametrisation}.  

\subsection{Causal (ir)regularity}\label{sec:causal regularity}

Discussed at length in~\cite{Asante:2021zzh,Asante:2021phx}, the Regge action can attain complex values when the imaginary parts of the dihedral angles at a face do not sum up to $\mp i2\pi$. This indicates the presence of causal irregularities in the form of a degenerate light cone structure. Two common examples for such configurations are the trouser singularity, where there are four instead of two light cones at the crotch of the trouser, or the yarmulke singularity, where there is no light cone at the tip. Choosing \enquote{$+$} in the convention of Lorentzian angles, the amplitude $\e^{i\Sreg}$ is exponentially suppressed for trouser configurations and exponentially enhanced for yarmulke configurations and vice versa for the choice \enquote{$-$}~\cite{Asante:2021zzh}.  

These two examples of causality violations are characterised by Lorentzian dihedral angles, located at the hinges of the cellular complex, and therefore referred to as hinge causality violations. It is furthermore proposed in~\cite{Asante:2021phx}, that the concept of hinge causality can be generalised to higher dimensions, referred to as edge and vertex causality. In the following, we study the three types of causality conditions for the present model and for the different signatures the sub-cells can take.

\paragraph{Hinge causality.} As indicated for the examples of trouser and yarmulke configurations, hinge causality violations are characterised by a non-vanishing imaginary part of the Lorentzian deficit angle. Each summand of $\mp i\frac{\pi}{2}$ for a dihedral angle counts one light cone crossing. Thus, there are exactly four light rays at a hinge if the imaginary parts of the dihedral angles sum up to $\mp i2\pi$. In the present setting, there are two different types of hinges to consider, squares and trapezoids, both of which are shared by four $4$-frusta. We consider these two types separately in the following.

At a square in slice $n$ there are four dihedral angles located, being two times the angle $\varphi_{nn-1}$ and two times the angle $\varphi_{nn+1}$. Following from Eqs.~\eqref{eq:phi nn+1} and~\eqref{eq:phi n+1n}, the explicit form of these dihedral angles depends on the signature of the associated $3$-frusta. Modulo time inversion, there exist five distinct possibilities for the signature and orientation of the two $3$-frusta attached to the $n$-th slice. In Tab.~\ref{tab:causality violations}, we summarise all the possibilities where \enquote{tl} indicates a timelike frustum and \enquote{sl $\uparrow (\downarrow)$} denotes a spacelike $3$-frustum with future (past) pointing $4$-dimensional normal vector.  

\begin{table}[h!]
    \centering
    \begin{tabular}{c|c c c c c}
         $3$-frustum in slab $(n-1,n)$ & tl & sl $\uparrow$ & sl $\downarrow$ & sl $\uparrow$ & sl $\uparrow$ \\[7pt]
         \hline\\[-5pt]
         $3$-frustum in slab $(n,n+1)$ & tl & tl            & tl              & sl $\uparrow$ & sl $\downarrow$\\[7pt]
         \hline\\[-5pt]
         regular causality & $\checkmark$ & X & X & $\checkmark$ & X
    \end{tabular}
    \caption{Causal regularity $(\checkmark)$ or irregularity (X) for the dihedral angle located at a square in the $n$-th slice where \enquote{tl} and \enquote{sl} indicate a timelike, respectively spacelike $3$-frustum and where $\uparrow$ and $\downarrow$ indicate the relative time orientation of the $4$-dimensional normal vector. Since the choice of time orientation is arbitrary, the regularity of time reversed configurations is the same.}
    \label{tab:causality violations}
\end{table}

We find that the two causally regular configurations are given when either the $3$-frusta of the two slices are timelike or if they are spacelike with the same orientation. That is, causally irregular configurations are only obtained when $H_n^2<\frac{1}{4}(s_n-s_{n+1})^2$. Within a single given slice, where the dihedral angles $\varphi_{nn+1}$ are located at the boundary, the case of spacelike $3$-frusta yields imaginary contributions to the Regge action as there are either zero or four light cone crossings, depending on the orientation of the timelike normal vector. Only if another slice with spacelike $3$-frusta of the same orientation is glued, the imaginary parts can be compensated, yielding a real bulk deficit angle. 

At a trapezoid in the slab between slice $n$ and $n+1$, there are four dihedral angles which are the same as the they lie in the same slab. If the trapezoid is timelike, the dihedral angles are Euclidean and we find for the bulk deficit angle
\begin{equation}
\delta^{\mathrm{E}} = 2\pi - 4\theta_n^{\mathrm{E}},
\end{equation}
which is real by definition. If the trapezoid is spacelike, the dihedral angles are timelike and we have
\begin{align}
\begin{cases}
\mp i2\pi + 4\cosh^{-1}\left(\frac{(s_n-s_{n+1})^2}{(s_n-s_{n+1})^2-4H_n^2}\right)&\mathrm{if}\quad H_n^2<\frac{1}{4}(s_n-s_{n+1})^2,\\[10pt]
\pm i2\pi + 4\cosh^{-1}\left(\frac{(s_n-s_{n+1})^2}{4H_n^2-(s_n-s_{n+1})^2}\right) &\mathrm{if}\quad H_n^2 > \frac{1}{4}(s_n-s_{n+1})^2,
\end{cases}
\end{align}
where we used Eqs.~\eqref{eq:theta E} and~\eqref{eq:theta L} for the definition of the dihedral angles $\theta_n^{\mathrm{E}}$ and $\theta_n^{\mathrm{L}}$, respectively. As a result, a spacelike trapezoid necessarily induces causality violations. The sign of the imaginary part of the deficit angle depends on the signature of the associated $3$-frustum and the choice of convention. As noted above, this sign determines whether the Lorentzian amplitudes $\e^{i\Sreg}$ enhance or suppress trouser or yarmulke configurations, respectively.  

Together with the results of Table~\ref{tab:causality violations}, we observe that the configurations satisfying hinge causality are given by those which contain timelike trapezoids which already implies timelike $3$-frusta, that is $H_n^2>\frac{1}{2}(s_n-s_{n+1})^2$. Explicitly, the regular action is given by
\begin{equation}\label{eq:regular Regge action}
\begin{aligned}
S_{\mathrm{R}}^{(n)} & = 6(s_n^2-s_{n+1}^2)\sinh^{-1}\left(\frac{s_{n+1}-s_n}{\sqrt{4H_n^2-(s_n-s_{n+1})^2}}\right)\\[7pt]
& + 12\abs{k_n}\left[\frac{\pi}{2}-\cos^{-1}\left(\frac{(s_n-s_{n+1})^2}{4H_n^2-(s_n-s_{n+1})^2}\right)\right].
\end{aligned}
\end{equation}  
In the light of Fig.~\ref{fig:regions}, Sector II exhibits hinge causality violations at trapezoids while Sector I shows such violations at trapezoids and squares. Finally, notice that hinge causality is not sensitive to the signature of the edges connecting the $n$-th and the $(n+1)$-th slice, and therefore, both Sector III.1 and III.2 are considered causally regular at the hinges.  

\paragraph{Edge causality.} To set up the definition of edge causality, consider the plane $P$  orthogonal to a spacelike edge $e$\footnote{We do not consider timelike edges, since the corresponding orthogonal $3$-dimensional space has Euclidean signature and therefore no light cones. This is comparable to a timelike face which has a $2$-dimensional Euclidean space orthogonal to it, resulting necessarily in real Euclidean angles without imaginary parts.} located at the midpoint of the edge. Moreover, let $\Sigma = \cup_{\sigma \supset e}\sigma$ be the union of $4$-polytopes sharing $e$. The intersection $P\cup\Sigma$ results in a discrete $3$-dimensional Lorentzian space with spherical boundary, whose single bulk vertex is the midpoint of the edge $e$. Following the definition of~\cite{Asante:2021phx}, the causal structure at the edge is said to be regular, if there are exactly two, one past and one future pointing, light cones emanating from the midpoint of $e$. 

In the present setting there are two types of spacelike edges. One is given by edges that lie on the boundary of $3$-cubes. The other is given by edges connecting neighbouring slices, which are only spacelike if the inequality $H_n^2<\frac{3}{4}(s_n-s_{n+1})^2$ holds, as Tab.~\ref{fig:regions} shows. Due to the hypercubical combinatorics, an edge is shared by eight $4$-frusta. We discuss the two types of edge causality in the following.

\begin{figure}
    \begin{subfigure}{0.5\textwidth}
    \centering
    \includegraphics[width=0.9\linewidth]{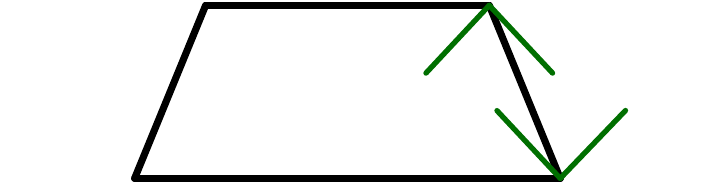}
    \end{subfigure}%
    \hfill
    \begin{subfigure}{0.5\textwidth}
    \centering
    \includegraphics[width=0.9\linewidth]{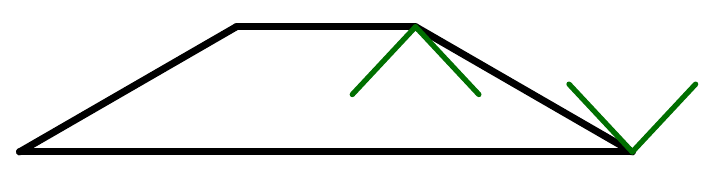}
    \end{subfigure}
    \caption{Light cones (green) emanating from vertices of a trapezoid with timelike (left) or spacelike (right) side edges. For a timelike edge, there is always one light ray inside the trapezoid at the upper and lower vertex. If the edge is spacelike, then the light cone of one vertex is fully contained inside the trapezoid while the light cone of the other vertex is fully outside the trapezoid. Causality violating configurations at spacelike squares, their edges or their vertices can be projected to two dimensions, yielding the figure on right-hand-side.}
    \label{fig:trapezoid_lightcone}
\end{figure}

Forming the intersection of the plane orthogonal to a spacelike edge $e$ inside a cube with a single $4$-frustum yields a $3$-frustum. Because of projecting out one dimension, the resulting $3$-frustum has a height $H_n$ and the squared length of its edges connecting two neighbouring slices is given by the squared height $_{(2)}h_n^2$ of the original (before projection) trapezoids,
\begin{equation}
_{(2)}h^2_n = \frac{(s_n-s_{n+1})^2}{2}-H_n^2.
\end{equation}
Then, the $3$-dimensional space obtained from the intersection of the plane with all the $4$-frusta is given by eight $3$-frusta glued together with the midpoint of edge $e$ as vertex in the middle. From this picture, we conclude that there are exactly two light cones if every single upper $3$-frustum contains a portion of the upper light cone, and every single lower $3$-frustum contains a portion of the lower light cone, each of which is emanating from the vertex $e$. This is the case if the edges connecting slices are timelike, i.e. if $_{(2)}h_m^2 < 0$ with $m\in\{n-1,n\}$. Then, by gluing the $3$-frusta along their faces, the pieces of light cones add up to two regular light cones. If one of the edges is instead spacelike, for instance if $_{(2)}h_n^2 > 0$, then each of the four $3$-frusta between slice $n$ and $n+1$ sharing the edge $e$ contains an entire or no light cone at all. In total, there are therefore either zero or four entire light cones in between the slice $n$ and $n+1$ in the projected $3$-frustum, necessarily inducing a degenerate light cone structure. Therefore, the causal structure at spacelike edges lying in $3$-cubes is regular if and only if the trapezoid is timelike, i.e. if $H_n^2 > \frac{1}{2}(s_n-s_{n+1})^2$. See Fig.~\ref{fig:trapezoid_lightcone} for an analogous $2$-dimensional depiction.

For a spacelike edge connecting slices $n$ and $n+1$, visualizing the orthogonal space and its intersection with the union of $4$-frusta is more involved. To study edge causality in this case, we explicitly compute the intersection of the plane with a single $4$-frustum. The total $3$-dimensional projected space is then obtained by gluing the eight intersections accordingly. A single $4$-frustum in $\R^{1,3}$ is defined by the sixteen vertices $\left(0,\pm\frac{s_n}{2},\pm\frac{s_n}{2},\pm\frac{s_n}{2}\right)$ and $\left(H_n,\pm\frac{s_{n+1}}{2},\pm\frac{s_{n+1}}{2},\pm\frac{s_{n+1}}{2}\right)$. Choosing two vertices of different slices connected by an edge, e.g. $\left(0,\frac{s_n}{2},\frac{s_n}{2},\frac{s_n}{2}\right)$ and $\left(H_n,\frac{s_{n+1}}{2},\frac{s_{n+1}}{2},\frac{s_{n+1}}{2}\right)$, its midpoint is given by
\begin{equation}
p = \left(0,\frac{s_n}{2},\frac{s_n}{2},\frac{s_n}{2}\right)+\frac{1}{2}\left(H_n,\frac{s_{n+1}-s_n}{2},\frac{s_{n+1}-s_n}{2},\frac{s_{n+1}-s_n}{2}\right).
\end{equation}
The plane $P$ orthogonal to the edge is defined as
\begin{equation}
P:-H_nx_0+\frac{s_{n+1}-s_n}{2}\sum_{i=1}^3x_i + d = 0,
\end{equation}
where $d(s_n,s_{n+1},H_n)$ is defined by demanding that $P$ contains the point $p$. Given the inequality $H_n^2<\frac{3}{4}(s_{n+1}-s_n)^2$ and assuming that $s_n> s_{n+1}$, the intersection of the plane $P$ with every edge of the $4$-frustum is non-zero only at the points
\begin{equation}
    q_i = -\frac{3(s_n-s_{n+1})^2-4H_n^2}{4(s_n-s_{n+1})}\vb*{e}_i+\frac{s_n}{2}\sum_{j}\vb*{e}_j,
\end{equation}
where $\vb*{e}_i$ are $4$-dimensional unit vectors in the $i$-th spatial direction. Together with the point $p$, these four points form a tetrahedron with three timelike faces and one spacelike face which is spanned by the points $q_i$. Each such $3$-dimensional Lorentzian building block contains no light cone.

\begin{figure}
    \centering
    \includegraphics[width=0.2\textwidth]{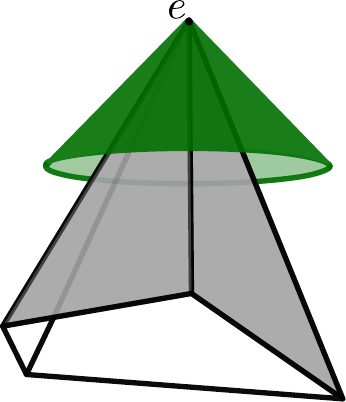}
    \caption{The space orthogonal to a spacelike edge $e$ connecting slices $n$ and $n+1$ with one dimension suppressed. Thus, in analogy, there are four flat timelike triangles instead of eight flat timelike tetrahedra depicted. The intersection of this piece-wise flat space with the light cone (in green) at the midpoint of edge $e$ is empty. Therefore, the causal structure at this edge is irregular.}
    \label{fig:edge_space}
\end{figure}

The total $3$-dimensional space orthogonal to the spacelike edge is obtained by gluing all of the eight tetrahedra along their timelike faces, while the spacelike faces form the spherical boundary. A visualisation of this space with one spatial dimension suppressed is given in Fig.~\ref{fig:edge_space}. As a result, this space does not contain a light cone at all and we conclude that edge causality is violated in every sector except Sector III.2. 

\paragraph{Vertex causality.} At any vertex $v$ of the cellular complex, consider the union of all $4$-dimensional building blocks that contain that vertex. Then, the causal structure is said to be regular if the intersection of the light cone at $v$ with the boundary of the union are two disconnected spheres~\cite{Asante:2021phx}. 

A vertex $v$ in a slice $n$ is shared by sixteen $4$-frusta, eight lying in the slab $(n-1,n)$ and eight lying in the slab $(n,n+1)$. If, for instance, the edge connecting slices $n$ and $n+1$ is spacelike, i.e. if $H_n^2 <\frac{3}{4}(s_n-s_{n+1})^2$, then each $4$-frustum of the slab $(n,n+1)$ at $v$ contains either an entire or no light cone at all, depending on the sign of $s_n-s_{n+1}$. The analogous two-dimensional situation is depicted on the right-hand-side of Fig.~\ref{fig:trapezoid_lightcone}. In this case, the vertex $v$ has either eight or zero upper light cones. The same applies to the case of a spacelike edge lying inside the slab $(n-1,n)$. Only if the edges are timelike, the light cone structure at the vertex $v$ is non-degenerate. Hence, vertex causality is violated in every sector except Sector III.2.

Summarizing, Sector I and II exhibit causality violating configurations at vertices, edges and hinges of the discrete Lorentzian spacetime. In Sector III.1 edge and vertex causality are violated which however remains unnoticed by the Regge action, which is only sensitive to hinge causality. Although the present setting is strongly simplified by symmetry, the results on causality violations suggest that higher-dimensional causality conditions imply lower-dimensional ones, e.g. vertex causality implies hinge causality. It is conceivable that this arises due to the dimensional reduction that results from the projections onto orthogonal subspaces of $k$-dimensional cells (e.g. edges, $k=1$ or hinges, $k=2$). However, we point out that the study of different causality violations and their interrelation in a more general setting remains an interesting question for future research.

\subsection{Wick rotation}\label{sec:Wick rotation} The symmetry reduced setting, and in particular the presence of the variable $H$ which is associated to the lapse function $N$, allows to define a Wick rotation between the Lorentzian theory, characterised by Eq.~\eqref{eq:slice Regge action} and the Euclidean theory, defined by the action~\cite{Bahr:2017bn}
\begin{equation}
\begin{aligned}
S_{\mathrm{R}}^{\mathrm{E}} =& 6(s_n^2-s_{n+1}^2)\left[\frac{\pi}{2}-\cos^{-1}\left(\frac{s_{n+1}-s_n}{\sqrt{4H_{n,\mathrm{E}}^2+(s_n-s_{n+1})^2}}\right)\right]\\[7pt]
+&
12\abs{k_{n,\mathrm{E}}}\left[\frac{\pi}{2}-\cos^{-1}\left(\frac{(s_n-s_{n+1})^2}{4H_{n,\mathrm{E}}^2+(s_n-s_{n+1})^2}\right)\right].
\end{aligned}
\end{equation}
In the context of Lorentzian Regge calculus, Ref.~\cite{Asante:2021phx} significantly advanced these studies by introducing a complexification of the Regge action that allows for a rigorous formulation of a Wick rotation. 

Analogous to continuum studies~\cite{Feldbrugge:2017kzv}, the Wick rotation is performed by complexifying the lapse function, represented in the discrete setting by the squared height $H^2\rightarrow R^2\e^{i\alpha}$, where $\alpha$ is the rotation angle and $R^2$ is the modulus. Consequently, the area of trapezoids as well as the dihedral angles are complexified, denoted by $k_n(\alpha),\theta^\pm_n(\alpha)$ and $\varphi^\pm_{nn+1}(\alpha)$. The resulting Regge action can be analytically continued, which crucially depends on the structure of branch cuts. This in turn depends on the causal sector, i.e. on the difference of spacelike edge length $s_n-s_{n+1}$ relative to $R_n$. 

In general, the complexified action for a single $4$-frustum, which we refer to as $\mathcal{S}_{\mathrm{R}}$, is of the form
\begin{equation}
\mathcal{S}_{\mathrm{R}}(\alpha) = 6(s_n^2-s_{n+1}^2)\left(\frac{\pi}{2}\pm\varphi_{nn+1}^{\pm}(\alpha)\right)+12k_n(\alpha)\left(\frac{\pi}{2}\pm \theta_n^{\pm}(\alpha)\right),
\end{equation}
which can be analytically continued to Wick rotation angles $\alpha\in (-2\pi,2\pi]$ depending on its branch cuts. As elaborated in~\cite{Asante:2021phx}, to which we refer to for further details, causally regular configurations (i.e. $2 R_n^2 > (s_n-s_{n+1})^2$) show the following behaviour under Wick rotation
\begin{equation}\label{eq:S limits 1}
\mathcal{S}_{\mathrm{III}}(\alpha)\longrightarrow
\begin{cases}
\; -S_{\mathrm{R}}^{\mathrm{E}},&\quad\text{for $\alpha = 0$}\, ,\\
-iS_{\mathrm{R}}^{\mathrm{L}},&\quad\text{for $\alpha = -\pi$}\, ,\\
\quad\, S_{\mathrm{R}}^{\mathrm{E}},&\quad\text{for $\alpha = 2\pi$}\, ,\\
\quad\! iS_{\mathrm{R}}^{\mathrm{L}},&\quad\text{for $\alpha = \pi$}\, ,
\end{cases}
\end{equation}
where we remind the reader that the regular Lorentzian action does not depend on the choice of sign for Lorentzian angles.

For causality violating terms in Sector I, $4R_n^2 < (s_n-s_{n+1})^2$, and II, $(s_n-s_{n+1})^2< 4R_n^2 < 2(s_n-s_{n+1})^2$, the Wick rotation is sensitive to the direction from which the values $\alpha =\pm\pi$ are approached. In particular, it is shown in~\cite{Asante:2021phx} that
\begin{equation}\label{eq:S limits 2}
\mathcal{S}_{\mathrm{I,II}}(\alpha)\longrightarrow
\begin{cases}
-iS_{\mathrm{R}}^{\mathrm{L,+}},&\quad\text{for $\alpha \rightarrow -\pi\uparrow$}\, ,\\
-iS_{\mathrm{R}}^{\mathrm{L,-}},&\quad\text{for $\alpha \rightarrow -\pi\downarrow$}\, ,\\
\quad\! iS_{\mathrm{R}}^{\mathrm{L,+}},&\quad\text{for $\alpha \rightarrow \pi\uparrow$}\, ,\\
\quad\! iS_{\mathrm{R}}^{\mathrm{L,-}},&\quad\text{for $\alpha \rightarrow \pi\downarrow$}\, ,
\end{cases}
\end{equation}
where the Lorentzian actions $S_{\mathrm{R}}^{\mathrm{L},\pm}$ now explicitly depend on the choice of sign. Approaching the Euclidean sector at $\alpha = 0,2\pi$, we notice that there exist two analytic extensions per sector, $\mathcal{S}_{\mathrm{I,II}}$ and $\mathcal{S}_{\mathrm{I,II}}'$, agreeing with $\mathcal{S}_{\mathrm{III}}$ in the regions $\alpha\in(-\pi,\pi)$ and $\alpha\in(-2\pi,\pi)\cup(\pi,2\pi]$, respectively. Consequently, 
\begin{equation}\label{eq:S limits 3}
\mathcal{S}_{\mathrm{I,II}}(0) = -S_{\mathrm{R}}^\mathrm{E},\qquad \mathcal{S}_{\mathrm{I,II}}'(2\pi) = S_{\mathrm{R}}^{\mathrm{E}}.
\end{equation}
In the other limiting cases, additional terms enter that depend on the number of light cone crossings. As explained above, Sector I shows causality violations at trapezoids and squares while Sector II is irregular only at trapezoids. As a result, the remaining limiting values of $\mathcal{S}_{\mathrm{I,II}}$ and $\mathcal{S}_{\mathrm{I,II}}'$ are given by
\begin{subequations}\label{eq:S limits 4}
\begin{align}
\mathcal{S}_{\mathrm{I}}'(0) &=  - S_{\mathrm{R}}^{\mathrm{E}}+2\pi\cdot 12\abs{k_n} - 2\pi  \cdot 6(s_n^2-s_{n+1}^2)\, , \\\mathcal{S}_{\mathrm{II}}'(0) &=  - S_{\mathrm{R}}^{\mathrm{E}}-4\pi\cdot 12\abs{k_n}\, , \\
\mathcal{S}_{\mathrm{I}}(2\pi) &= \;\;S_{\mathrm{R}}^{\mathrm{E}} -2\pi\cdot 12\abs{k_n} + 2\pi\cdot 6(s_n^2-s_{n+1}^2)\, ,\\
\mathcal{S}_{\mathrm{II}}(2\pi) &= \;\;S_{\mathrm{R}}^{\mathrm{E}} +4\pi\cdot 12\abs{k_n}\, .
\end{align}
\end{subequations}

In summary, the Lorentzian and Euclidean Regge actions are related via the Wick rotation procedure described in this section, where causality violating terms in the Lorentzian sector will also lead to changes in the Euclidean sector. This direct connection hinges on the existence of a foliation into spacelike hypersurfaces which, similar to the setting in CDT, allows for a global Wick rotation in a single parameter. It is nevertheless remarkable that the Lorentzian action obtained in Sec.~\ref{sec:Lorentzian Regge action for spatially flat cosmology} is the result of an ad hoc construction and still directly connected to the Euclidean action derived in~\cite{Bahr:2017bn} from the semi-classics of an underlying quantum geometric model. Although this result suggests that there may exist a symmetry restricted causally extended Lorentzian spin foam model whose semi-classical limit leads to the Lorentzian Regge action of spatially flat cosmology, the ultimate answer to this questions remains open. That is because the semi-classical limit of full Lorentzian spin foam models, being either the CH-extension of the EPRL model~\cite{Conrady:2010vx,Conrady:2010kc} or the causal completion of the Barrett-Crane model~\cite{Jercher:2022mky}, is in general not well understood and subject to active research, see~\cite{Simao:2021qno}.

\section{A minimally coupled massless scalar field} \label{sec:Minimally coupled scalar field}

In this section, we introduce a minimally coupled massless scalar field to the cosmological model precedingly developed. Starting from a continuum perspective a scalar field, being a $0$-form, is discretised by placing it either on primary or dual vertices of the cellular complex~\cite{Desbrun:2005ug}. Since we assume spatial homogeneity also for the scalar field, it is for our purposes advantageous to discretise the scalar field on primal vertices, $\phi(t)\rightarrow \phi_n$, which therefore attains the index of the slice. 

\subsection{Discrete scalar field action}

The continuum action for a free massless scalar field, minimally coupled to spatially flat cosmology, given in Eq.~\eqref{eq:scalar field action cont}, is quadratic in fields and derivatives and exhibits translation and reflection symmetry. Following~\cite{Hamber:2009wl,Hamber:1993gn}, we translate this action to the discrete setting. The scalar field action for a single $4$-frustum between slice $n$ and $n+1$ is then given by
\begin{equation}\label{eq:slice scalar field action}
S_\phi^{(n)}[\phi_n,\phi_{n+1},s_n,s_{n+1},H_n] = w_n(\phi_{n+1}-\phi_n)^2.
\end{equation}
Clearly, this action exhibits translation and reflection symmetry and is quadratic in the fields. Here, $w_n(s_n,s_{n+1},H_n)$ is a geometrical coefficient that plays the role of the continuum factor $\frac{a^3}{2N}$ in the discrete and scales like squared length. Therefore, we choose
\begin{equation}\label{eq:wn ansatz}
w_n = \frac{1}{2H_n^2}\left(\frac{s_n+s_{n+1}}{2}\right)^3H_n = \frac{(s_n+s_{n+1})^3}{16H_n},
\end{equation}
which yields the correct continuum limit of the coupled gravity and matter system, as we show in Sec.~\ref{sec:Linearisation, deparametrisation and continuum limit}.\footnote{Notice, that the continuum limit of the scalar field equations, which we consider in Sec.~\ref{sec:scalar field cont limit}, does only fix the function $w_n$ up to a function $f$ of the ratio $\eta_n \equiv\frac{s_{n+1}-s_n}{H_n}$. Instead, demanding the correct continuum limit of the fully coupled gravity and matter system fixes $w_n$ in leading order of $\eta_n$ to Eq.~\eqref{eq:wn ansatz}.} The first expression of Eq.~\eqref{eq:wn ansatz} is closest to the continuum, where $1/H_n^2$ encodes the discrete squared time derivative. The lengths $(s_n+s_{n+1})/2$ and the height $H_n$ characterise the volume of a $4$-dimensional cuboid which can be understood as the discretisation of the $4$-volume element $\sqrt{-g}$.  

On an extended $2$-complex with $L$ spatial vertices and spatial $M+1$ slices, the total action for $\phi$ is given by
\begin{equation}\label{eq:scalar field action}
S_\phi = L\sum_{n = 0}^{M-1}w_n(\phi_n-\phi_{n+1})^2,
\end{equation}
showing the same slicing structure as the Regge action in Eq.~\eqref{eq:Regge action}. This simplifies the equations of motion significantly, as the equation on the $n$-th slice does only depend on the dynamical quantities of the neighbouring slices $n-1$ and $n+1$.

\subsection{Equations of motion}

In this section, we consider the equations of motion of the scalar field on the homogeneous background with $M+1$ slices and fixed initial and final values $\phi_0$ and $\phi_M$. To that end, we perform the variation of $S_\phi$ in Eq.~\eqref{eq:scalar field action}, yielding the system of equations
\begin{equation}\label{eq:discrete scalar field equation}
\phi_n(w_{n-1}+w_n)-(\phi_{n-1}w_{n-1}+\phi_{n+1}w_n) = 0,\qquad n \in \{1,...,M-1\},
\end{equation}
which can be solved for $\phi_n$, yielding
\begin{equation}\label{eq:phi_n}
\phi_n = \frac{\phi_{n+1}\frac{1}{w_{n-1}}+\phi_{n-1}\frac{1}{w_n}}{\frac{1}{w_{n-1}}+\frac{1}{w_n}}.   
\end{equation}
Recursively, this system of equations can be solved for a single scalar field $\phi_n$ as a function of the boundary data $\phi_0,\phi_M$ and the geometric quantities $\{s_n,H_n\}$. For the first recursion step, we insert the equation for $\phi_{n-1}$ into the equation for $\phi_{n}$, yielding
\begin{equation}
    \phi_{n} = \frac{\phi_{n+1}\left(\frac{1}{w_{n-2}}+\frac{1}{w_{n-1}}\right)+\phi_{n-2}\left(\frac{1}{w_n}\right)}{\sum\limits_{m=n-2}^n\frac{1}{w_m}}
\end{equation}
Repeating these steps to \enquote{earlier} and \enquote{later} slices until one reaches the initial and final slices $0$ and $M$, respectively, the scalar field solutions is given by
\begin{equation}\label{eq:phi_n sol}
\phi_n = \frac{\phi_M\mathcal{W}_0^{n-1}+\phi_0\mathcal{W}_n^{M-1}}{\mathcal{W}_0^{M-1}},
\end{equation} 
where we introduced the symbols
\begin{equation}
\mathcal{W}_{n_1}^{n_2} = \sum_{m =n_1}^{n_2}\frac{1}{w_m},
\end{equation}
which explicitly depend on the geometric quantities $\{H_n,s_n\}$. Clearly, scalar field solutions for boundary conditions $\phi_0 = \phi_M$ are constant, i.e. $\phi_n = \phi_0$ for all $n\in\{0,...,M\}$. 

To check for monotonicity, it is instructive to re-express the scalar field solutions in the following form
\begin{equation}
\phi_{n+1}-\phi_n = \frac{\frac{1}{w_n}}{\mathcal{W}_0^{M-1}}(\phi_M-\phi_0).
\end{equation} 
Since each $w_m$ is positive, following from Eq.~\eqref{eq:wn ansatz}, the scalar field evolves monotonically, the sign of which depends on the sign of $\phi_M-\phi_0$, i.e.
\begin{equation}
\mathrm{sgn}\left(\phi_{n+1}-\phi_n\right) = \mathrm{sgn}\left(\phi_M-\phi_0\right),\qquad \forall n\in\{0,...,N-1\}.
\end{equation}
For the later purpose of deparametrising the system with respect to the scalar field, this is one of the crucial properties.

\subsection{Continuum time limit}\label{sec:scalar field cont limit}

The simple form of the scalar field equations allows for a straightforward definition of a map from discrete to continuum variables. Since the scalar field equation is coupled non-trivially to the geometry via the factors $w_n$ that enter Eq.~\eqref{eq:discrete scalar field equation}, such a map needs to be defined for the matter variables $\phi_n$ as well as for the geometrical variables $\{s_n\}$ and $\{H_n\}$.  

As $H_n$ labels the distance of slices $n$ and $n+1$, it is associated to a finite proper time difference and thus directly related to the lapse. Therefore, in the continuum time limit where $H_n$ becomes infinitesimal, we identify $H_n\rightarrow \dd{\tau} = N\dd{t}$. Since $n\in\{0,...,M\}$ labels the spatial slices which, in continuum cosmology, correspond to slices of constant time, we propose the following map for the variables $s_n$ and $\phi_n$~\cite{Bahr:2017bn}
\begin{align}
s_n\rightarrow a(t),&\qquad s_{n\pm 1}\rightarrow a(t) \pm \dot{a}(t)\dd{t} +\frac{1}{2}\left(\ddot{a}(t)-\frac{\dot{N}}{N}\dot{a}(t)\right)\dd{t}^2,\label{eq:s cont limit}\\[7pt]
\phi_n \rightarrow \phi(t),&\qquad \phi_{n\pm 1} \rightarrow \phi(t) \pm \dot{\phi}(t)\dd{t}+\frac{1}{2}\left(\ddot{\phi}(t)-\frac{\dot{N}}{N}\dot{\phi}(t)\right)\dd{t}^2,\label{eq:phi cont limit}
\end{align}
where $a(t)$ is the scale factor, $\phi(t)$ the continuum scalar field, $N(t)$ is the lapse and dot denotes a derivative with respect to time $t$.

Applying this limit to the scalar field equation, we find at lowest order in $\dd{t}$
\begin{equation}
3\dot{a} a^2  \dot{\phi} + a^3\ddot{\phi} -\frac{\dot{N}}{N}a^3\dot{\phi} = 0,
\end{equation}
which can be brought to the familiar form
\begin{equation}
\dv{}{t}\left(\frac{a^3}{N}\dot{\phi}\right) = 0,
\end{equation}
corresponding to the continuum scalar field equation. 

\subsection{Total action of the coupled system}

We close the section on the scalar field by providing a formula for the total action, governing the dynamics of the coupled system with gravity and matter. Importantly both, the Regge action in Eq.~\eqref{eq:Regge action} and the scalar field action in Eq.~\eqref{eq:scalar field action} split into a sum over slices. Consequently, the total action is simply given by
\begin{equation}\label{eq:total action}
S_{\mathrm{tot}} = L\sum_{n = 0}^{M-1}S_{\mathrm{tot}}^{(n)} = L\sum_{n = 0}^{M-1}\left(\frac{1}{8\pi \GN}S_{\mathrm{R}}^{(n)}+\lambda S_\phi^{(n)}\right).
\end{equation}
We introduced the factor of Newton's coupling $\GN$ and a parameter $\lambda$ in the sum of the two actions in order to: (i) account for the correct dimensions of the two terms and (ii) have a parameter that controls the strength and sign of the matter term. 

\paragraph{Wick rotation of total action.} Finally, we comment on the possibility of defining a Wick rotation for the total action, analogous to what we have discussed in Sec.~\ref{sec:Wick rotation}. A complexification $H^2\rightarrow R^2\e^{i\alpha}$ of the height variable induces a complexification of the single scalar field action in Eq.~\eqref{eq:slice scalar field action}. An analytic continuation thereof is straightforwardly defined as,
\begin{equation}
\mathcal{S}_\phi(\alpha) = \e^{-i\alpha/2}\frac{(s_n-s_{n+1})^2}{R}\left(\phi_n-\phi_{n+1}\right)^2,
\end{equation}
which is extended to $\alpha\in(-2\pi,2\pi]$.\footnote{Notice, that in the presence of a mass term, entering as $\sim V_n m^2\phi_n^2\sim H_n m^2\phi_n^2$, the Wick rotation leads to a  relative minus sign between the derivative and the mass term. This is comparable to the continuum situation, where a mass term would enter Eq.~\eqref{eq:scalar field action cont} with an additional factor of $N^2$.} The Euclidean and Lorentzian sectors are then obtained in the limits
\begin{equation}
\mathcal{S}_\phi(0) = S_{\phi}^{(n)},\qquad \mathcal{S}_{\phi}(\pm\pi) = \mp iS_\phi^{(n)},\qquad \mathcal{S}_\phi(2\pi) = -S_{\phi}^{(n)}.
\end{equation}
Thus, the analytic continuation of gravitational and matter actions need to be subtracted to yield the correct limits, i.e.
\begin{equation}\label{eq:total complex action}
\mathcal{S}_{\mathrm{tot}}(\alpha) = \frac{1}{8\pi G_{\mathrm{N}}}\mathcal{S}_{\mathrm{R}}(\alpha)-\lambda\mathcal{S}_\phi(\alpha).
\end{equation}
For causality violating terms, where the Wick rotation is non-trivial, we discuss the classical equations of motion of the total system in Sec.~\ref{sec:Classical equations and causality violations}.

\section{Regge equations and deparametrisation}\label{sec:Regge equations and deparametrisation} 

We study the classical equations of motion derived from the Regge action of the total system of geometry and matter, defined by the action given in Eq.~\eqref{eq:total action}.

In the homogeneous setting with $M+1$ slices, there are in total $2M-1$ geometric bulk variables, being $M-1$ spatial edge lengths $\{s_n\}$ and $M$ lengths $\{l_n\}$ associated to edges that connect neighbouring spatial slices. Following the introduction of Regge calculus in the beginning of Sec.~\ref{sec:Spatially flat cosmology in Regge calculus}, the Regge equations are obtained via a variation\footnote{As noted in~\cite{Bahr:2017bn}, varying with respect to the homogeneous variables $(s_n,H_n)$ corresponds to a global variation. A local variation, where the homogeneity is imposed afterwards is another possibility, for details of which we refer to~\cite{Brewin1999}.} with respect to the edge length, which in this case yields a coupled system of $2M-1$ transcendental equations, obtained from the variations
\begin{equation}
\frac{\partial S_{\mathrm{tot}}}{\partial l_n}\overset{!}{=} 0,\qquad \frac{\partial S_{\mathrm{tot}}}{\partial s_m}\overset{!}{=} 0,
\end{equation}
where $n\in\{0,...,M-1\}$ and $m\in\{1,...,M-1\}$. Since it is convenient to work with height variables $\{H_n\}$ instead of edge lengths $\{l_n\}$, the first set of Regge equations can be re-expressed as
\begin{equation}
\pdv{S_{\mathrm{tot}}}{H_n}\pdv{H_n}{l_n} = 0\quad\Rightarrow\quad \pdv{S_{\mathrm{tot}}}{H_n} = 0,
\end{equation}
where the Jacobian between length and height variables is invertible as long as the edges are not lightlike. 

Due to the slicing structure of the total Regge action, the equations of motion for variables on the $n$-th slice only depend on the variables of the neighbouring slices. This holds independently of the causal structure. For causally regular configurations with Regge action as in Eq.~\eqref{eq:regular Regge action}, the general form of the Regge equations is given by
\begin{subequations}\label{eq:general cosmological Regge equations}
\begin{align}
\pdv{S_{\mathrm{tot}}}{H_n} =& \pdv{\abs{k_n}}{H_n}\left[\frac{\pi}{2}-\cos^{-1}\left(\frac{\Delta s_n^2}{4H_n^2-\Delta s_n^2}\right)\right]-\lambda \pi \GN\frac{(s_n+s_{n+1})^3}{24 H_n^2}\Delta\phi_n^2 = 0,\label{eq:general cosmological Regge equations 1}\\[10pt]
\pdv{S_{\mathrm{tot}}}{s_n} =& \pdv{\abs{k_{n-1}}}{s_n}\left[\frac{\pi}{2}-\cos^{-1}\left(\frac{\Delta s_{n-1}^2}{4H_{n-1}^2-\Delta s_{n-1}^2}\right)\right] + \pdv{\abs{k_{n}}}{s_n}\left[\frac{\pi}{2}-\cos^{-1}\left(\frac{\Delta s_n^2}{4H_n^2-\Delta s_n^2}\right)\right]\nonumber \\[7pt]
+& s_n\left[\sinh^{-1}\left(\frac{\Delta s_{n-1}}{\sqrt{4H_{n_1}^2-\Delta s_{n-1}^2}}\right)-\sinh^{-1}\left(\frac{\Delta s_n}{\sqrt{4H_n^2-\Delta s_n^2}}\right)\right]+\label{eq:general cosmological Regge equations 2} \\[7pt]
+& \lambda\pi\GN\left[\frac{(s_{n-1}+s_n)^2}{8H_{n-1}}\Delta\phi_{n-1}^2+\frac{(s_n+s_{n+1})^2}{8H_n}\Delta\phi_n^2\right] = 0,\nonumber
\end{align}
\end{subequations}
where we imposed the Schl\"{a}fli identity and used $\Delta s_n = s_n-s_{n+1}$ and $\Delta\phi_n = \phi_n-\phi_{n+1}$ as a short-hand notation. Notice that the overall factor $L$ of the number of cubes per spatial slice drops out of the equations. The derivatives of the trapezoid area with respect to $H_n$ and $s_n$ are respectively given by
\begin{equation}\label{eq:area derivatives}
\pdv{\abs{k_n}}{H_n} = \frac{H_n(s_n+s_{n+1})}{\sqrt{H_n^2-\frac{\Delta s_n^2}{2}}},\qquad \pdv{\abs{k_n}}{s_n} = \frac{H_n^2-s_n\Delta s_n}{2\sqrt{H_n^2-\frac{\Delta s_n^2}{2}}}
\end{equation}

In the following three sections, we restrict our attention to configurations that satisfy hinge causality, i.e. $H_n^2>\frac{1}{2}(s_{n}-s_{n+1})^2$ holds for every slab. Only for these configurations the number of real equations is the same as the number of variables. A discussion of the Regge equations for hinge causality violating terms is given in Sec.~\ref{sec:Classical equations and causality violations}.

\subsection{Vacuum Regge equations}\label{sec:Vacuum Regge equation}

Deficit angles are defined in terms of inverse trigonometric and hyperbolic functions. As a consequence, the Regge equations, summarised in Eqs.~\eqref{eq:general cosmological Regge equations}, are transcendental and therefore cannot be solved analytically. Also a numerical treatment of the equations is complicated since the initial conditions for solving algorithms are required to be close to the actual solutions for a reliable convergence.

In order to gain a first grasp of their structure, we consider the simplified setting of a vanishing matter contribution, either by setting $\lambda = 0$ or by choosing boundary conditions such that $\Delta\phi_n=0$ everywhere. Furthermore, we consider first the case of zero bulk spatial slices to deduce from it the general case of $M+1$ slices. 

Given two spatial slices and fixed boundary values $s_0,s_1$, the only dynamical bulk variable is $H_0$ governed by Eq.~\eqref{eq:general cosmological Regge equations 1} with $\lambda  = 0$. The derivative of the trapezoid area in Eq.~\eqref{eq:area derivatives} never vanishes for finite boundary data. Furthermore the deficit angle $\theta_n$ is monotonic in $\Delta s/H_0$ with its only root given when $\Delta s = 0$. Therefore, one does not find a local extremum of the Regge action for any $H_0\in\R$ if $\Delta s \neq 0$. The only configuration for which the derivative of $\Sreg$ vanishes is when the action itself vanishes. This is the case for boundary data corresponding to flat geometries, i.e. when $s_0 = s_1$. In this case, all the deficit angles vanish for arbitrary values of $H_0$,
\begin{equation}
\eval{S_{\mathrm{R}}}_{s_0 = s_1} = 0,\qquad \forall\: H_0\in\R.
\end{equation}
We provide an interpretation of this result at the end of the next paragraph.

Following the same argument, the equations of motion for the $M$ bulk variables $\{H_n\}$ can only be satisfied if the deficit angles vanish, i.e. if $\Delta s_n = 0$ for all $n$. Then the Regge action vanishes and thus becomes independent of the height variables. The length variables $s_n$ associated to slices in the bulk are dynamical and governed by Eq.~\eqref{eq:general cosmological Regge equations 2}. Also these equations are satisfied on vanishing deficit angles, being therefore consistent with the equations for $H_n$. As a result, imposing the equations of motion on every slice, we obtain an extended globally flat cubulation, the height variables of which become arbitrary.

Studying the vacuum Regge equations of the symmetry reduced model yields two insights. First, the only solution that is compatible with the absence of matter is the flat solution for which the Regge action vanishes. This is consistent with the dynamics of the continuum, where the scale factor $a$ is constant in the spatially flat vacuum case. The second insight is, that for flat configurations, i.e. $s_n = s_{n+1}$ for all $n$, the height variables $\{H_n\}$ become arbitrary. The emergence of such a symmetry is common for flat solutions in 4d Regge calculus~\cite{Rocek:1981ama,Rocek:1982tj} and signifies the restoration of diffeomorphism symmetry in the discrete~\cite{Dittrich:2008pw,Bahr:2009qc}. Due to the restricted setting and global evolution, this symmetry amounts to moving spacelike slices (instead of individual vertices), see also the discussion in~\cite{Bahr:2015gxa}. This is analogous to the arbitrariness of the lapse function in the continuum.  

\subsection{Regge equations with matter} 

In this and the following section, we look at the full system defined by the total action in Eq.~\eqref{eq:total action}, focusing on the causally regular Sector III.

\begin{figure}
    \begin{subfigure}{0.5\textwidth}
    \centering
    \includegraphics[width=0.9\linewidth]{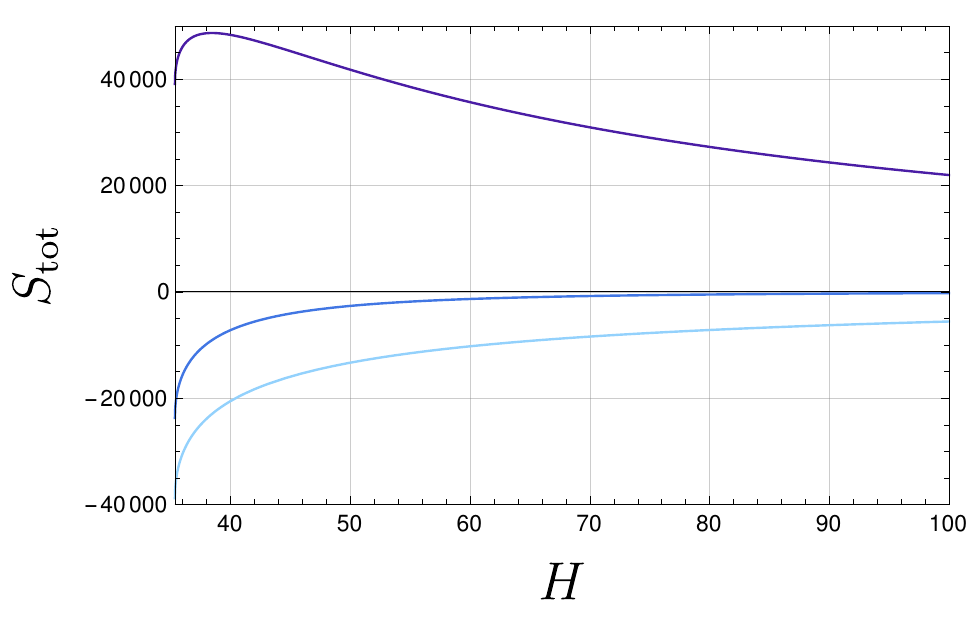}
    \end{subfigure}%
    \begin{subfigure}{0.5\textwidth}
    \centering
    \includegraphics[width=0.9\linewidth]{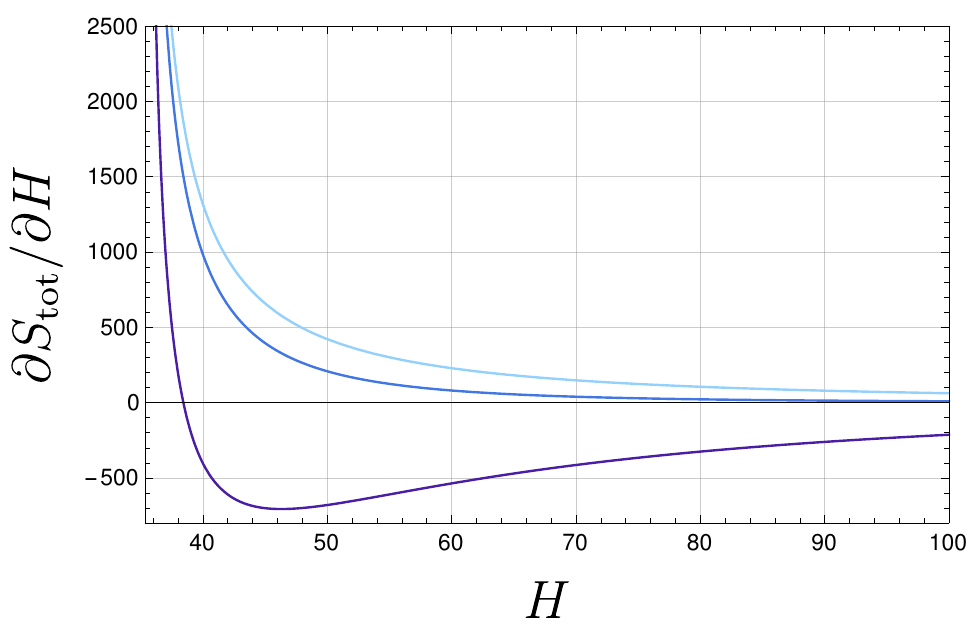}
    \end{subfigure}
    \caption{Total Regge action (left) and its derivative (right) for a single slice with boundary data $s_0 = 50, s_1 = 100$ and scalar field differences $\Delta\phi_*-2$ (light blue), $\Delta\phi_*$ and $\Delta\phi_*+2$ (dark blue). Here, we set $\lambda = 8\pi\GN = 1$ such that the limiting value of Eq.~\eqref{eq:ineq for existence of sols} is given by $\Delta\phi_* = 2\sqrt{\frac{2}{3}}$. For $\Delta\phi <\Delta\phi_*$ there is no solution to the equation of motion, while for $\Delta\phi>\Delta\phi_*$ there exists a solution.}
    \label{fig:matter_regge}
\end{figure}

If $M=1$, where the only dynamical variable is $H_0$, there exist solutions of the equation
\begin{equation}\label{eq:non-perturbative H eq}
6\frac{H_0(s_0+s_1)}{\sqrt{H_0^2-\frac{(s_1-s_0)^2}{2}}}\left[\frac{\pi}{2}-\cos^{-1}\left(\frac{(s_1-s_0)^2}{4H_0^2-(s_1-s_0)^2}\right)\right]-\lambda 8\pi\GN\frac{(s_0+s_1)^3}{16H_0^2}(\phi_1-\phi_0)^2 = 0
\end{equation}
for non-trivial boundary data $s_0\neq s_1$ and $\phi_0\neq \phi_1$ and provided that $\lambda >0$. A plot for fixed spatial edge length and varying field values is shown in Fig.~\ref{fig:matter_regge}. From these plots, we observe that there is a condition on the boundary data for the existence of solutions. This condition is given by
\begin{equation}\label{eq:ineq for existence of sols}
\left(\frac{2}{s_0+s_1}\right)^2\left(\frac{s_1-s_0}{\phi_1-\phi_0}\right)^2 < \lambda\frac{4\pi\GN}{3},
\end{equation}
and we show in Sec.~\ref{sec:Linearisation, deparametrisation and continuum limit} that it arises in a small deficit angle expansion. From this inequality, we extract that a large change of spatial edge length needs to be accompanied by large scalar field field differences such that solutions are allowed. Notice that configurations violating this inequality only admit Euclidean solutions, which is a consequence of the sign flip in the argument of the inverse cosine function entering the deficit angles of the Euclidean theory. 

For more than two spatial slices, say $M+1$, non-trivial equations for bulk spatial length $\{s_n\}$ and scalar field values $\{\phi_n\}$ must be satisfied. As for the case of $M=1$, the equations for the height variables $\{H_n\}$ exhibit solutions, provided that the inequality of Eq.~\eqref{eq:ineq for existence of sols} holds on every slice. Notice that these conditions now also affect bulk spatial lengths and scalar field values. That is, the equations of motion for the $\{H_n\}$ constrain the equations of bulk spatial edge lengths and scalar fields. 

Assuming that the equation $\partial S_{\mathrm{tot}}/\partial H_n = 0$ is satisfied, i.e. Eq.~\eqref{eq:general cosmological Regge equations 1} holds, we extract
\begin{equation}
\frac{\pi}{2}-\cos^{-1}\left(\frac{\Delta s_n^2}{4H_n^2-\Delta s_n^2}\right) = -\left(\pdv{\abs{k_n}}{H_n}\right)^{-1}\pdv{S_{\phi}^{(n)}}{H_n}.
\end{equation}
Implicitly, the height variables $\{H_n\}$ are therefore given as functions of the spatial edge length and the scalar field variables, i.e. $H_n = H_n(\{s_m\},\{\phi_m\})$. As a result, the equation for spatial edge length $s_n$, explicitly given in Eq.~\eqref{eq:general cosmological Regge equations 2}, reduces to
\begin{equation}\label{eq:sinh relation}
\sinh^{-1}\left(\frac{s_n-s_{n-1}}{\sqrt{4H_{n-1}^2-(s_{n}-s_{n-1})^2}}\right) = \sinh^{-1}\left(\frac{s_{n+1}-s_n}{\sqrt{4H_n^2-(s_{n+1}-s_n)^2}}\right).
\end{equation}
We used the equation
\begin{equation}
\pdv{\abs{k_n}}{s_n}\left(\pdv{\abs{k_n}}{H_n}\right)^{-1}\pdv{S_\phi^{(n)}}{H_n} = \pdv{S_\phi^{(n)}}{s_n},
\end{equation}
which holds by chain rule and applies if the functions are invertible. This is the case if we exclude lightlike trapezoids for which $k_n$ vanishes, i.e. we demand that $H_n^2 \neq \frac{1}{2}\Delta s_n^2$.

Finally, Eq.~\eqref{eq:sinh relation} implies that
\begin{equation}\label{eq:non-perturbative s eq}
\frac{s_{n+1}-s_n}{H_n} = \frac{s_n-s_{n-1}}{H_{n-1}}.
\end{equation}
This means that relative to the $4$-height, the differences of spatial edge lengths within a $4$-frustum remain constant along the entire discrete spacetime.

Solving for $s_n$, we find
\begin{equation}
s_n = \frac{s_{n-1} H_n+s_{n+1}H_n}{H_{n-1}+H_{n-1}},
\end{equation}
where we notice the structural similarities compared to the scalar field equation~\eqref{eq:phi_n}. Given boundary data $(s_0,s_M)$, and introducing the sum of $4$-heights between slice $m_1$ and $m_2$,
\begin{equation}
\mathcal{H}_{m_1}^{m_2} = \sum_{m=m_1}^{m_2}H_m
\end{equation}
the $n$-th spatial edge length is given by
\begin{equation}
s_n = \frac{s_M\mathcal{H}_0^{n-1}+s_0\mathcal{H}_n^{M-1}}{\mathcal{H}_0^{M-1}}.
\end{equation}
Notice that the scalar field dependence of the spatial edge length $s_n$ enters via the height variables $\{H_n\}$.

With these formulas, the differences of neighbouring spatial length, say on slice $n+1$ and $n$ is expressed as
\begin{equation}
\frac{s_{n+1}-s_n}{H_n} = \frac{s_M-s_0}{\mathcal{H}_0^{M-1}}.
\end{equation}
That is, the length difference of spacelike edges relative to the height of a given $4$-frustum is the same as the length difference of the boundary spacelike edges, $(s_M-s_0)$ relative to the total height $\mathcal{H}_0^{M-1}$. Importantly, from this equation it follows that the spatial edge length evolve monotonically, with the sign determined by relation of $s_M$ and $s_0$, i.e.
\begin{equation}\label{eq:s monotonic}
\mathrm{sgn}(s_{n+1}-s_n) = \mathrm{sgn}(s_M-s_0),\qquad \forall n\in\{0,...,M-1\}.
\end{equation}
Consequently, the boundary data selects either the contracting or expanding branch, similar to what is found in the continuum, see Appendix~\ref{sec:Relational cosmology in the continuum}.

\subsection{Linearisation, deparametrisation and continuum limit}\label{sec:Linearisation, deparametrisation and continuum limit}

The discussion on the solutions of the Regge equations has been mostly qualitative so far. To obtain a more quantitative understanding, we consider in this section the expansion of the Regge equations around small deficit angles of both types, i.e. $\varphi_{nn+1}\ll 1$ and $\theta_n\ll 1$. This procedure will reduce the transcendental equations to polynomial ones. 

Both types of deficit angles are a function of
\begin{equation}\label{eq:eta}
\eta_n = \frac{s_{n+1}-s_n}{H_n}.
\end{equation}
Clearly, expanding in small $\eta_n$ yields an expansion in small deficit angles with the flat case given by $s_n = s_{n+1}$, i.e. $\eta_n = 0$. Notice also that in order to remain within the causally regular sector, $\eta_n$ must obey the bound $\eta_n^2 < 2$. An expansion of the equation $\partial S/\partial H_n = 0$ around $\eta_n = 0$ yields
\begin{equation}\label{eq:1st CRE expansion}
6(s_n+s_{n+1})\left(\frac{\eta_n^2}{4}+\frac{\eta_n^4}{8}+\mathcal{O}(\eta_n^6)\right)-\lambda 8\pi\GN\frac{(s_n+s_{n+1})^3}{16 H_n^2}(\phi_{n+1}-\phi_n)^2 = 0.
\end{equation}
As a consistency check, we notice that cutting the equation at vanishing order of $\eta_n$ yields the flat solution where $\Delta s_n = 0$ and $\Delta \phi_n = 0$ for all slices $n$.

At first non-vanishing order, $\eta_n$ enters Eq.~\eqref{eq:1st CRE expansion} quadratically and the dependence on the height variable $H_n$ drops out, yielding
\begin{equation}\label{eq:discrete relational Friedmann}
\left(\frac{2}{s_n+s_{n+1}}\right)^2\left(\frac{s_{n+1}-s_n}{\phi_{n+1}-\phi_n}\right)^2 = \lambda\frac{4\pi\GN}{3},
\end{equation}
which corresponds exactly yo the limiting case of the inequality~\eqref{eq:ineq for existence of sols}. This equation is \textit{relational} in the sense that it only involves the spatial edge length and the scalar field values, independent of the height of the frusta. In fact, one can show that the $H_n$ are undetermined at this level of expansion by the whole system of equations, similar to the vacuum case of Sec.~\ref{sec:Vacuum Regge equation}. It is at this level of expansion, where one can define a continuum time limit, which we discuss in the following paragraph.

\paragraph{Continuum time limit.} Defining a continuum limit for the general Regge action is a challenging task. However, in the symmetry reduced model we are considering here, the system is effectively one-dimensional with non-trivial dynamics in temporal direction. Schematically speaking, the continuum limit corresponds to infinitesimally small but many time steps, at each of which the deficit angles are small~\cite{Bahr:2017bn,Brewin1999}. Consequently, this limit is two-fold. First, as performed for the scalar field in Sec.~\ref{sec:scalar field cont limit}, the height of a $4$-frustum is interpreted as finite proper time which, in the continuum limit, is mapped to $H_n\rightarrow \dd{\tau} = N\dd{t}$. Spatial edge lengths $\{s_n\}$ and scalar field values $\{\phi_n\}$ are mapped according to Eqs.~\eqref{eq:s cont limit} and~\eqref{eq:phi cont limit}, respectively. The second limit needed to recover the Friedmann equations is that of small deficit angles which has been performed above as an expansion in the parameter $\eta_n$. Following this prescription, we start with Eq.~\eqref{eq:discrete relational Friedmann} and find at vanishing order of $\dd{t}$
\begin{equation}
\left(\frac{\dot{a}}{a}\right)^2 = \lambda\frac{4\pi \GN}{3}\dot{\phi}^2,
\end{equation}
corresponding indeed the continuum Friedmann equation.

Given the already relational form of Eq.~\eqref{eq:discrete relational Friedmann}, introducing an artificial time parameter $t$ with lapse $N$ is actually not necessary. Instead, we consider the continuum limit  $\phi_{n+1}-\phi_n\rightarrow \dd{\phi}$ (which implicitly assumes the scalar field differences to be approximately constant along the slices) and understand the spatial length variables as functions of the relational clock $\phi$, yielding
\begin{equation}
s_n\rightarrow a(\phi),\qquad s_{n\pm 1}\rightarrow a(\phi)\pm a'(\phi)\dd{\phi}+\frac{1}{2}a''(\phi)\dd{\phi}^2,
\end{equation}
where prime denotes the derivative with respect to the scalar field. Then, at vanishing order of $\dd{\phi}$, Eq.~\eqref{eq:discrete relational Friedmann} becomes
\begin{equation}
\left(\frac{a'}{a}\right)^2 = \lambda\frac{4\pi \GN}{3},
\end{equation}
corresponding to the relational continuum Friedmann equation. 

Since in the continuum the lapse function corresponds to a gauge redundancy, the set of equations after deparametrisation should be in fact smaller than in the discrete, where the variable $H_n$ is in fact dynamical. This reduction of independent equations can be seen from the non-perturbative (i.e. at all order in $\eta_n$) equation~\eqref{eq:non-perturbative s eq}. Clearly, in the continuum time limit as prescribed by Eqs.~\eqref{eq:s cont limit} and~\eqref{eq:phi cont limit}, Equation~\eqref{eq:non-perturbative s eq} takes the form $\dot{a} = \dot{a}$ which becomes trivially satisfied.

In the discrete, Eq.~\eqref{eq:discrete relational Friedmann} can be solved explicitly for one of the edge lengths, say $s_{n+1}$. Solving the polynomial of second degree, we obtain
\begin{equation}\label{eq:s sol linearised}
s_{n+1}^+ = \frac{1+\sqrt{\lambda\frac{\pi\GN}{3}(\phi_{n+1}-\phi_n)^2}}{\abs{1-\sqrt{\lambda\frac{\pi\GN}{3}(\phi_{n+1}-\phi_n)^2}}}s_n,\qquad s_{n+1}^- = \frac{\abs{1-\sqrt{\lambda\frac{\pi\GN}{3}(\phi_{n+1}-\phi_n)^2}}}{1+\sqrt{\lambda\frac{\pi\GN}{3}(\phi_{n+1}-\phi_n)^2}}s_n,
\end{equation}
corresponding to the expanding and contracting solutions, respectively. Since this relation can be derived for any slice, and since Eq.~\eqref{eq:s monotonic} dictates global monotonicity, we find a relation between distant slices by iteration,
\begin{equation}
s_M^{\pm} = \prod_{n=0}^{M-1}\kappa^{\pm}_n s_0,
\end{equation}
where $\kappa^{\pm}_n$ are the defining coefficients of Eq.~\eqref{eq:s sol linearised}. Approximating the scalar field differences as $\phi_{n+1}-\phi_n = (\phi_M-\phi_0)/M$ and using the limit definition of the exponential function, $\lim_{n\rightarrow\infty}(1+x/n)^n = \e^x$, we find
\begin{equation}
\lim\limits_{M\rightarrow\infty} s_M^{\pm} = s_0\exp\left(\pm\sqrt{\lambda\frac{4\pi\GN}{3}}(\phi_M-\phi_0)\right),
\end{equation}
which corresponds indeed to the relational continuum solution of Eq.~\eqref{eq:cont relational Friedmann}.

\paragraph{Higher orders.} Including higher orders of the parameter $\eta_n$ to the full system of equations increases in general the degree of the resulting polynomial equations. Already at fourth order in $\eta_n$, the equation $\partial S/\partial H_n = 0$ shows qualitative differences compared to the lowest order case. In particular, the equation becomes explicitly dependent on the height variable $H_n$, for which one can solve explicitly
\begin{equation}
H_n^2 = \frac{\frac{1}{2}\Delta s_n^4}{\lambda \frac{4\pi\GN}{3}\left(\frac{s_n+s_{n+1}}{2}\right)^2\Delta\phi_n^2-\Delta s_n^2}.
\end{equation}
Here, we see explicitly that the inequality~\eqref{eq:ineq for existence of sols} is required to hold in order to yield a real Lorentzian height $H_n\in\R$. From the plots of Fig.~\ref{fig:matter_regge}, we see that this inequality is a non-perturbative property that will therefore hold at all orders. Furthermore, these plots suggest that if the inequality is satisfied, there exists only a single solution. Consequently, it is expected that for higher orders of the polynomial equation (admitting in principle many solutions), there exists at most one viable solution for the height variable.\footnote{We thank an anonymous referee for pointing this out.}

\paragraph{Causality violations and continuum limit.} Only causally regular terms admit an expansion in terms of small parameters $\eta_n$ defined in Eq.~\eqref{eq:eta}. That is because causality violating terms are given for $\eta_n^2 > 2$ and therefore, an expansion for $\eta_n\ll 1$ is not justified. Following the construction of the continuum limit, such configurations therefore do not admit a limit of $H_n\rightarrow N\dd{t}$ and simultaneously taking the deficit angles to be small. In the present symmetry reduced setting, this suggests that causality violations are a feature of the discrete theory that are not visible in an appropriate continuum limit. A more general result on the interplay of discreteness and causal structure remains however an open question.

\subsection{Classical equations and causality violations}\label{sec:Classical equations and causality violations}

In the preceding three sections, we focused on the Regge equations of (hinge) causally regular configurations satisfying the inequality $H_n^2 >\frac{1}{2}(s_n-s_{n+1})^2$. There, the Regge action is real such that the number of real equations equals the number of real dynamical variables. Hence the analysis has been performed in a standard way. However, if hinge causality violations are present, the Regge action is complex, therefore leading to complex equations of motion for a priori real dynamical variables. We discuss in the following how such types of systems can be studied along the lines of~\cite{Asante:2021phx}.

Complex actions for gravity have been studied in the context of the Euclidean gravity path integral~\cite{Louko:1995jw,Gibbons:1976ue,Gibbons:1978ac,Witten:2021nzp}, continuum Lorentzian quantum cosmology~\cite{Feldbrugge:2017kzv}, semi-classical analyses of Lorentzian spin foams~\cite{Han:2020npv,Han:2023cen,Han:2021kll} as well as Lorentzian effective spin foams~\cite{Asante:2021phx,Asante:2021zzh}. Therein, the path integral over real metric configurations is replaced by a complex line integral over a contour $C$ in the space of complex metrics, formally written as
\begin{equation}
    Z = \int_C\mathcal{D}g\e^{\frac{1}{\hbar}\mathcal{S}[g]},
\end{equation}
where $\mathcal{S}[g]$ is an analytic functional of the complex metric $g$. Choosing the contour to be a Lefshetz-thimble~\cite{Lefshetz1975,Vassiliev2002}, i.e. a path of steepest descent that passes through one or more critical points $g_0$, the convergence of the integral improves significantly. For explicit examples in the setting of discrete and continuum quantum cosmology, see~\cite{Asante:2021phx} and~\cite{Feldbrugge:2017kzv}, respectively. In an asymptotic expansion $\hbar\rightarrow 0$, the saddle points $g_0$ dominate the path integral. Therefore, as in standard quantum mechanics, the saddle points of the action can be interpreted as the classical (however complex) configurations. 

Applied to the present setting of complex Regge actions, we conclude that a complexification of the variables is necessary to obtain an equal number of equations and dynamical variables. Given two slices, i.e. $M = 1$, the only dynamical variable is the squared height $H^2 = R^2\e^{i\alpha}$ for which the corresponding analytic actions $\mathcal{S}(R,\alpha)$ have been discussed in Sec.~\ref{sec:Wick rotation}. For $M > 1$, the spatial edge lengths in the bulk also become dynamical, necessitating further complexification and analytic continuation. In the following, we restrict ourselves to the case of $M = 1$.

In Sector III, the total complex action $\mathcal{S}_{\mathrm{III}}^{\mathrm{tot}}$ has saddle points either on Lorentzian data, $\alpha = \pm\pi$, or on Euclidean data, $\alpha = 0,2\pi$, depending on the boundary data. More precisely, if Eq.~\eqref{eq:ineq for existence of sols} is satisfied, there are Lorentzian solutions. If Eq.~\eqref{eq:ineq for existence of sols} is violated, then there are Euclidean solutions given at $\alpha=0,2\pi$. Both solutions require that $\lambda > 0$.

\begin{figure}
    \centering
    \includegraphics[width=0.7\textwidth]{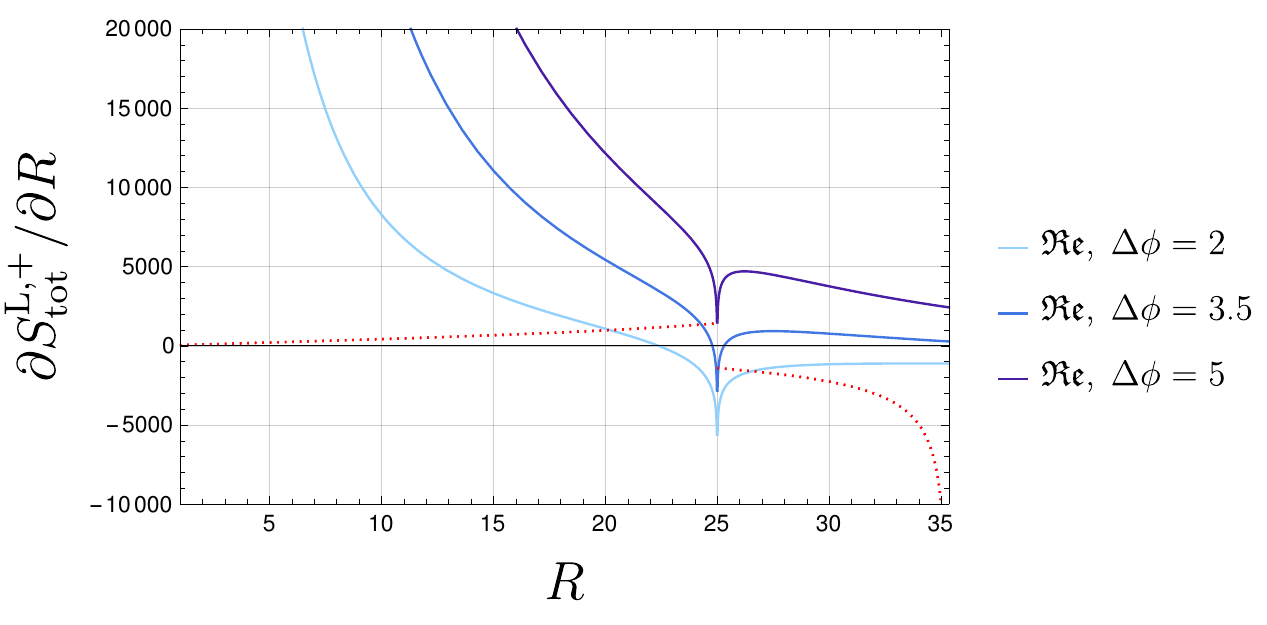}
    \caption{Real (solid blue) and imaginary (dashed red) parts of the derivative of the total Regge action in Sectors I and II with the choice of convention \enquote{$+$}. Boundary data is given by $s_0 = 50, s_1 = 100$ and parameter values $\lambda = -1$ and $8\pi\GN =1$ are chosen. Although the real part shows saddle points for certain values of $\Delta\phi$, the imaginary part is nowhere vanishing.}
    \label{fig:dReggeIandII}
\end{figure}

Sectors I and II include hinge causality violations and are described by the complex actions $\mathcal{S}^{\mathrm{tot}}_{\mathrm{I}}$ and $\mathcal{S}^{\mathrm{tot}}_{\mathrm{II}}$, respectively. Their limiting values $\alpha\rightarrow\pm \pi$ and $\alpha\rightarrow 0,2\pi$ are summarised in Eqs.~\eqref{eq:S limits 2}--\eqref{eq:S limits 4}. From these limiting values, consider for instance the Lorentzian action $S_{\mathrm{tot}}^{\mathrm{L},+}$. Its derivative with respect to the radial variable $R$ is depicted in Fig.~\ref{fig:dReggeIandII}. We observe that for certain boundary data $(s_0,\phi_0)$ and $(s_1,\phi_1)$, the real part (drawn in solid blue) exhibits a saddle point with respect to the radial variable $R$ given that $\lambda <0$. However, the imaginary part (drawn in dashed red) is never zero, in particular not at the zeroes of the real part.\footnote{This argument holds also for other choices of sign which only reflect the imaginary part at the $R$-axis.} As a consequence, the causality violating sectors do not exhibit saddle points on the Lorentzian lines $(R,\alpha) = (R,\pm\pi)$. We therefore conclude that there are no classical solutions for the Lorentzian action with causality violations.

Similarly, there are no saddle points on Euclidean lines $(R,0)$ and $(R,2\pi)$ for causality violating data, where $R^2<\frac{1}{4}\Delta s^2$ in Sector I and $\frac{1}{4}\Delta s^2 < R^2<\frac{1}{2}\Delta s^2$ in Sector II. Expressions of the action on these lines are given in Eqs.~\eqref{eq:S limits 4}. 

These results suggest, that there is no classical correspondence for causality violating terms in the Euclidean or Lorentzian sector. Consequently, such configurations are treated purely quantum, and are only taken into account in a path integral. We cannot exclude the existence of saddle points of $\mathcal{S}^{\mathrm{tot}}_{\mathrm{I}}$ and $\mathcal{S}^{\mathrm{tot}}_{\mathrm{II}}$ in the fully complex regime, i.e. when $\alpha$ is not given by $0,2\pi$ or $\pm\pi$. However, the results of~\cite{Asante:2021phx} suggest that also for other angles $\alpha$, no saddle points exist. Even if existent, a geometric interpretation of such configurations beyond spatial topology change or the examples discussed in~\cite{Witten:2021nzp} would pose new challenges, going beyond the scope of this work. Saddle points of complex actions might also require the violation of energy conditions, on which we comment below.

\paragraph{A comment on energy conditions.} The Regge equation for a single slice exhibits a causally regular solution, either Lorentzian or Euclidean, if the coupling parameter $\lambda$ of the matter term in Eq.~\eqref{eq:scalar field action} is positive. As discussed above, Sectors I and II possibly exhibit complex solutions that require $\lambda$ to be negative. In this case, an analogy to the continuum could be drawn, where the sign of $\lambda$ corresponds to the sign of the energy density, given by the $(00)$-component of the energy-momentum tensor. The case of $\lambda > 0$ yields an energy-momentum tensor that satisfies the strong energy condition~\cite{HawkingBook}, which has been shown in~\cite{Tipler:1977eb} to guarantee a causally regular structure without singularities. If the strong energy condition is violated, which in this case would correspond to $\lambda  < 0$, causality violations cannot be excluded. These continuum results are exemplary for the intimate relation between matter and the causal structure and possibly translate to the present discrete setting. Further investigation in this direction is left open as an interesting future avenue.

\section{Discussion and conclusion}\label{sec:Discussion and conclusion}

The purpose of this article was to develop and study the classical discrete theory for spatially flat, homogeneous and isotropic cosmology within the framework of Lorentzian Regge calculus. As matter content of the universe, we investigated a minimally coupled massless free scalar field which is commonly employed in quantum cosmology to serve as a relational clock. A firm understanding of the kinematics and dynamics of the coupled system serves as a foundation for future investigations of the symmetry restricted path integral in the spirit of effective spin foams~\cite{Asante:2020qpa,Asante:2020iwm,Asante:2021zzh,Dittrich:2023rcr}. 

Starting from a triangulation that allows the split into triangulated spacelike hypersurfaces we imposed spatial homogeneity and isotropy. Together with the requirement of spatial flatness, we obtained a discretisation consisting of $4$-dimensional frusta that are internally triangulated in a flat way. In Secs.~\ref{sec:Lorentzian frustum}--\ref{sec:Lorentzian Regge action for spatially flat cosmology}, we derived the corresponding Lorentzian Regge action and its geometrical ingredients as a function of the edge length of cubes and the $4$-heights of frusta. We examined in Sec.~\ref{sec:causal regularity} the different sectors of the theory, summarised in Fig.~\ref{fig:regions}, with regard to their causality properties. Notably, causality violations, as introduced in~\cite{Asante:2021phx}, are present if edges, trapezoids or $3$-frusta between slices are spacelike. A Wick rotation to the Euclidean sector can be defined by complexifying the height variables and analytically continuing the action. As we discussed in Sec.~\ref{sec:Wick rotation} this requires particular care in the presence of hinge causality violations. 

We introduced a minimally coupled massless free scalar field to the discrete cosmological model in Sec.~\ref{sec:Minimally coupled scalar field}, studied its equations of motion and showed that solutions evolve monotonically as dictated by the boundary data. Moreover, it is straightforward to define a continuum limit of the scalar field equations, which we showed in Sec.~\ref{sec:scalar field cont limit} to agree with the continuum equations given in Appendix~\ref{sec:Relational cosmology in the continuum}. 

In accordance with the continuum, the vacuum Regge equations exhibit only static solutions where the spacelike geometric data remains constant along discrete temporal evolution. Indeterminacy of the $4$-height is in this case reminiscent of a restoration of diffeomorphism symmetry which, in the cosmological setting, is reflected in the lapse function $N$ being a gauge redundancy. In the presence of a non-vanishing scalar field, the causally regular Regge equations for the height variables exhibit solutions if the inequality~\eqref{eq:ineq for existence of sols} is satisfied. Furthermore, we extracted from the full equations of motion that the spacelike edge lengths evolve monotonically, constituting a contracting or an expanding branch. In Sec.~\ref{sec:Linearisation, deparametrisation and continuum limit} we studied a small deficit angle expansion of the transcendental Regge equations. At lowest non-vanishing order, the dependence on the height variable drops out similar to the vacuum case. This marks the level at which the system deparametrises and the evolution of spatial geometric data is described in a relational sense with respect to the scalar field. Moreover, a continuum limit at this level of truncation yields the correct relational Friedmann equations. Taking higher orders into account, the height variables become dynamical, obstructing a deparametrisation of the system. Lastly in Sec.~\ref{sec:Classical equations and causality violations} we discussed the existence and characterisation of classical solutions in the presence of causality violations.
\\
\\
\indent From our investigation of this cosmological model, we can draw a few tentative conclusions. We observe that causality violations generically appear if the building blocks connecting spatial slices are spacelike. This applies in particular to the situation in which the 3-frusta and trapezoids are entirely spacelike. For these sectors of the theory no solutions exist, at least for real length variables, and the entirely timelike sector appears preferred both in vacuum and in the presence of a massless scalar field. This casts doubts on the viability to describe cosmology in such a symmetry reduced setting with entirely spacelike 3d building blocks, as prescribed by the Lorentzian EPRL model. However, this is not conclusive from our analysis, as the symmetry reduction used here is highly restrictive. It is conceivable that these issues can be circumvented by deviating from homogeneity and isotropy and that these notions could reappear (effectively) after coarse graining or averaging. Nevertheless, recent works in group field cosmology~\cite{Jercher:2023nxa,Jercher:2023kfr} suggest that timelike polyhedra need to be included to obtain GR-like perturbations.

The simple cosmological model described in this article provides a possibility to study the refinement limit in the quantum theory, similar to the Euclidean considerations~\cite{Bahr:2018gwf}. On the one hand, we can study bulk refinements by adding more time steps to the triangulation. On the other hand, we can refine the boundary. While the former adds additional degrees of freedom, the latter does not due to the strong symmetry restrictions. Thus, this setting can give first insights into the definition of a continuum limit. Once the symmetries, i.e. homogeneity and isotropy, are eventually lifted, more interesting scenarios can be explored. For example, one can conceive a triangulation in which different spatial slices have a different number of faces and edges. Thus, under evolution degrees of freedom are added or removed with the spin foam acting as a dynamical embedding map relating states in different Hilbert spaces. See~\cite{Dittrich:2009fb,Dittrich:2011ke} and~\cite{Dittrich:2013xwa,Hohn:2014uvt,Hohn:2014rba} for the exploration of these ideas in Regge calculus and path integrals respectively. Such a procedure might help to address the question whether an effective cosmological dynamics emerges from the full theory via coarse graining.

We close this work by listing intriguing questions that wait to be addressed by the path integral quantisation of the model developed here.

\begin{itemize}
    \item \textit{Transfer matrix method:} The slicing structure of the action possibly allows to cast the system in a transfer matrix formalism and study the properties of the transition amplitudes similar to Causal Dynamical Triangulations~\cite{Ambjorn:2012pp,Ambjorn:2014mra}.
    \item \textit{Boundary states and semi-classics:} It will be interesting to study the definition of suitable boundary states that are peaked on the scale factor and under which conditions semi-classical physics are recovered from the quantum theory. Our present analysis of the classical dynamics will be crucial in this task.
    \item \textit{Acceleration operators:} Evaluating the effective Lorentzian spin foam sum is computationally challenging even in a strongly symmetry restricted model as the one developed here. That is due to the oscillatory nature of the effective spin foam amplitudes of the form $\e^{iS_\mathrm{R}}$. The recently re-discovered acceleration operators might prove highly useful to improve convergence~\cite{Dittrich:2023rcr}. Including causally irregular configurations and comparing to the spatially spherical model~\cite{Dittrich:2023rcr} will be interesting to study as well.
    \item \textit{Quantum bounce:} Studying the existence of a quantum bounce in effective spin foams is expected to be insightful for the mechanism of such bouncing scenarios. In particular, this might answer if semi-classical amplitudes together with discrete area spectra are sufficient for a singularity resolution.
    \item \textit{Deparametrisation:} Beyond the vacuum theory, the interplay between geometry and the massless scalar field is intriguing. In particular, the question arises whether the theory can be deparametrised and under which conditions this is possible. We have already observed that classically, reparametrisation invariance gets restored in a continuum limit, in which the deficit angles are small. This is consistent, as we expect a restoration of continuum symmetries in a suitable refinement limit, see e.g. the recent review~\cite{Asante:2022dnj}. Still, conceptual questions remain, e.g. with respect to which observable to deparametrise and whether this idea might be spoiled by fluctuations of matter and gravity.
\end{itemize}

\paragraph{Acknowledgements}

The authors are grateful to Seth Asante, Markus Schr\"{o}fl, Luca Marchetti, Daniele Oriti and Steffen Gielen for helpful comments and discussions. AFJ and SSt gratefully acknowledge support by the Deutsche Forschungsgemeinschaft (DFG, German Research Foundation) project number 422809950. AFJ kindly acknowledges support from the MCQST via the seed funding Aost 862983-4 and from the Deutsche Forschungsgemeinschaft (DFG) under Grant No 406116891 within the Research Training Group RTG 2522/1.

\appendix

\section{Lorentzian dihedral angles of $4$-frusta}\label{sec:Lorentzian dihedral angles of 4-frusta}

In this appendix we derive the dihedral angles for the Lorentzian $4$-frustum following the conventions of~\cite{Sorkin:2019llw,Asante:2021zzh}. In the boundary of a $4$-frustum there are three distinct $3$-dimensional building blocks, being the initial and final cube as well as the six boundary $3$-frusta. Due to homogeneity and isotropy, there are three distinct dihedral angles per slab $(n,n+1)$, being $\varphi_{nn+1}$, $\varphi_{n+1n}$ and $\theta_n$ located at the initial and final square and the trapezoids, respectively. 

We compute these angles by embedding a single $4$-frustum in flat Minkowski space $\R^{1,3}$ and determining the outward-pointing normal vectors of the $3$-dimensional building blocks. For initial and final cube, these are respectively given by $N_{s,\mp} = (\mp 1 ,0 ,0, 0)$. To obtain normal vectors of the $3$-frusta, we form the wedge product of three spanning vectors and act with the Hodge-star operator $*$. As a result, the six normal vectors are given by
\begin{equation}
N_{\pm,i} = \frac{1}{\sqrt{\abs{H_n^2-\left(\frac{s_n-s_{n+1}}{2}\right)^2}}}\left(\frac{s_n-s_{n+1}}{2}, \pm H_n\vb*{e}_i\right).
\end{equation}
The signature of $N_{\pm,i}$ is opposite of the signature of the $3$-frustum, i.e. $\eta(N_{\pm,i},N_{\pm,i}) = -1$ for the $3$-frustum to be spacelike and $\eta(N_{\pm,i},N_{\pm,i}) = 1$ for the $3$-frustum to be timelike. Notice that the time orientation for a timelike normal vector depends on whether $s_n < s_{n+1}$ or vice versa. 

Lorentzian dihedral angles are defined in terms of the Minkowski product of the normal vectors associated to the two polyhedra. For the three cases we consider here, these are given by
\begin{align}
    \eta(N_{\pm i},N_{\pm j}) &= -\frac{(s_n-s_{n+1})^2}{\abs{4H_n^2-(s_n-s_{n+1})^2}},\\[7pt]
    \eta(N_{\pm,i},N_{s,\mp}) &= \pm\frac{s_n-s_{n+1}}{\sqrt{\abs{4H_n^2-(s_n-s_{n+1})^2}}}.
\end{align}
The precise form of the Lorentzian angles $\varphi_{nn+1},\varphi_{n+1n}$ and $\theta_n$ as a function of the Minkowski products depend then on the signature of the $3$-polyhedra. Moreover, the signature of the trapezoid is decisive for the associated angle to be either Lorentzian or Euclidean. With the conventions of~\cite{Sorkin:2019llw,Asante:2021zzh}, the angles in Eqs.~\eqref{eq:phi nn+1}--\eqref{eq:theta E} follow. 

\section{Relational cosmology in the continuum}\label{sec:Relational cosmology in the continuum}

Here, we briefly summarise the equations of spatially flat cosmology with a minimally coupled massless scalar field, first in coordinates and then in a relational manner. The resulting formulae serve as a comparison for the continuum limit of the Regge equations, studied in Sec.~\ref{sec:Linearisation, deparametrisation and continuum limit}.

\subsection{Spatially flat cosmology and the lapse function}

As the starting point, we consider the spatially flat FLRW-metric, defined by the line element
\begin{equation}
\dd{s}^2 = g_{\mu\nu}\dd{x}^\mu\dd{x}^\nu = -N^2\dd{t}^2+a^2\dd{\vb*{x}}^2,
\end{equation}
where $N$ is the Lapse function, $a$ is the scale factor and $\dd{\vb*{x}}^2$ is the line element of flat $3$-dimensional Euclidean space. The corresponding symmetry reduced Einstein-Hilbert action is given by
\begin{equation}
S_{\mathrm{EH}}[a,N] = \frac{3}{8\pi\GN}\int\dd{t}\frac{a^3}{N}\left(\frac{\ddot{a}}{a}-\frac{\dot{N}}{N}\frac{\dot{a}}{a}+\left(\frac{\dot{a}}{a}\right)^2+N^2\frac{k}{a^2}\right).
\end{equation}
As matter content, we consider a minimally coupled massless scalar field, the action of which is given by
\begin{equation}\label{eq:scalar field action cont}
S_\phi[a,N,\phi] = \frac{\lambda}{2}\int\dd{t} \frac{a^3}{N}\dot{\phi}^2,
\end{equation}
yielding
\begin{equation}\label{eq:scalar field equation cont}
\dv{}{t}\left(\frac{a^3}{N}\dot{\phi}\right) = 0
\end{equation}
as equation of motion. 

Variation of the total action $S_{\mathrm{tot}} = S_{\mathrm{EH}}+S_\phi$ with respect to the lapse function $N$ yields the first Friedmann equation
\begin{equation}\label{eq:1st Friedmann cont}
\left(\frac{\dot{a}}{a}\right)^2+N^2\frac{k}{a^2} = \lambda\frac{4\pi \GN  }{3}\dot{\phi}^2,
\end{equation}
while the variation with respect to the scale factor $a$ yields the second Friedmann equation
\begin{equation}\label{eq:2nd Friedmann cont}
2\frac{\ddot{a}}{a}-2\frac{\dot{N}}{N}\frac{\dot{a}}{a}+\left(\frac{\dot{a}}{a}\right)^2+N^2\frac{k}{a^2}=-\lambda 4\pi\GN\dot{\phi}^2.
\end{equation}
Altogether, Eqs.~\eqref{eq:scalar field equation cont},~\eqref{eq:1st Friedmann cont} and~\eqref{eq:2nd Friedmann cont} form a set of two independent equations with the third equation being redundant. These two equations determine the dynamics of the physical variables $a$ and $\phi$. 

Notice that under a complexification of the lapse function $N\rightarrow iN_{\mathrm{E}}$ to Euclidean signature, the two Friedmann equations stay invariant if $k=0$, i.e. in the spatially flat case. 

\subsection{Deparametrisation in the continuum}

Following the relational strategy~\cite{Rovelli:1990ph,Dittrich:2005kc,Hoehn:2019fsy}, the dependence of the Friedmann equations on the lapse function $N$ and the unphysical time coordinate $t$ can be removed by deparametrising the system with respect to the scalar field $\phi$, thus serving as a relational clock. To that end, the scalar field is assumed to be on-shell, i.e.
\begin{equation}
\dv{\phi}{t} = \frac{cN}{a^3},
\end{equation}
where $c$ is the integration constant arising from solving Eq.~\eqref{eq:scalar field equation cont}. 

Then, the two Friedmann equations~\eqref{eq:1st Friedmann cont} and~\eqref{eq:2nd Friedmann cont} can be deparametrised, yielding
\begin{align}
\left(\frac{a'}{a}\right)^2+\frac{a^4}{c^2}k &=\lambda\frac{4\pi\GN}{3},\label{eq:rel Friedmann eq cont1}\\[7pt]
2\frac{a''}{a}-5\left(\frac{a'}{a}\right)^2+\frac{a^4}{c^2}k &= -\lambda 4\pi\GN.\label{eq:rel Friedmann eq cont2}
\end{align}
Notice that the second equation is a derivative of the first and hence dependent, because the scalar field equation has been imposed to deparametrise the system. As expected, the lapse function $N$ as well as the coordinate time $t$ vanished from the formalism completely with only the scale factor $a(\phi)$ as a function of the physical clock $\phi$ remaining.

For initial conditions $a(\phi_0) = a_0$ and in the standard case of $\lambda = 1$, the two solutions of this equation in the spatially flat case ($k=0$) are given by
\begin{equation}\label{eq:cont relational Friedmann}
a(\phi) = a_0\exp\left({\pm\sqrt{\frac{4\pi\GN}{3}}(\phi-\phi_0)}\right),
\end{equation}
where $\pm$ leads to the expanding, respectively the contracting branch of solutions.

\bibliographystyle{JHEP}
\bibliography{references}

\end{document}